\documentclass[%
preprint,
amsmath,amssymb,
aps,
prfluids,
groupedaddress,
]{revtex4-2}

\usepackage{graphicx}
\usepackage{dcolumn}
\usepackage{bm}
\usepackage{lipsum} 
\usepackage{subcaption}
\usepackage[colorlinks=true,linkcolor=blue]{hyperref}%
\usepackage{color}
\usepackage{CJKutf8}

\begin{document}

\begin{CJK*}{UTF8}{gbsn}
\title{Linear attention coupled Fourier neural operator for simulation of three-dimensional turbulence}
\author{Wenhui Peng (彭文辉)$^{1,2,3}$}
\author{Zelong Yuan (袁泽龙)$^{1,2}$}
\author{Zhijie Li (李志杰)$^{1,2}$}
\author{Jianchun Wang (王建春)$^{1,2}$}
\email{wangjc@sustech.edu.cn}
\affiliation{\small$^{1}$Department of Mechanics and Aerospace Engineering, Southern University of Science and Technology, Shenzhen 518055, China}
\affiliation{\small$^{2}$Guangdong-Hong Kong-Macao Joint Laboratory for Data-Driven Fluid Mechanics and Engineering Applications, Southern University of Science and Technology, Shenzhen 518055, China}
\affiliation{\small$^{3}$Department of Computer Engineering, Polytechnique Montreal, H3T1J4, Canada}

\date{\today}

\begin{abstract}

Modeling three-dimensional (3D) turbulence by neural networks is difficult because 3D turbulence is highly-nonlinear with high degrees of freedom and the corresponding simulation is memory-intensive. Recently, the attention mechanism has been shown as a promising approach to boost the performance of neural networks on turbulence simulation. However, the standard self-attention mechanism uses $O(n^2)$ time and space with respect to input dimension $n$, and such quadratic complexity has become the main bottleneck for attention to be applied on 3D turbulence simulation. In this work, we resolve this issue with the concept of linear attention network. The linear attention  approximates the standard attention by adding two linear projections, reducing the overall self-attention complexity from $O(n^2)$ to $O(n)$ in both time and space. The linear attention coupled Fourier neural operator (LAFNO) is developed for the simulation of 3D isotropic turbulence and free shear turbulence. Numerical simulations show that the linear attention mechanism provides 40\%
error reduction at the same level of computational cost, and LAFNO can accurately reconstruct a variety of statistics and instantaneous spatial structures of 3D turbulence. The linear attention method would be helpful for the improvement of neural network models of 3D nonlinear problems involving high-dimensional data in other scientific domains.

\end{abstract}

\maketitle
\end{CJK*}

\section{\label{intro}Introduction}
With the rising of deep learning techniques, neural networks (NNs) have been extensively explored to complement or accelerate the traditional computational fluid dynamics (CFD) modeling of turbulent flows \cite{brunton2020machine,duraisamy2019turbulence}. Applications of deep learning and machine learning techniques to CFD include approaches to the improvements of Reynolds averaged Navier–Stokes (RANS) and large eddy simulation (LES) methods. These efforts have mainly focused on using NNs to learn closures of Reynolds stress and subgrid-scale (SGS) stress and thus improve the accuracy of turbulence modeling \cite{maulik2019subgrid,ling2016reynolds,beck2019deep}.

Deep neural networks (DNNs) have achieved impressive performance in approximating the highly non-linear functions\cite{lecun2015deep}.
Guan et al. proposed the convolutional neural network model to predict the  SGS forcing
terms in two-dimensional decaying turbulence \cite{guan2022stable}.Yang et al. incorporated the vertically integrated thin-boundary-layer equations into the model inputs to enhance the extrapolation capabilities of neural networks for large-eddy-simulation wall modeling \cite{yang2019predictive}. Some recent works aim to approximate the entire Navier-Stokes equations by deep neural networks \cite{lusch2018deep,sirignano2018dgm,tang2021exploratory,sun2020neupde,kovachki2021neural,meng2022learning,linka2022bayesian,goswami2022deep,howard2022multifidelity}. Once trained, the ``black-box'' NN models can make inference within seconds on modern computers, thus can be extremely efficient compared with traditional CFD approaches \cite{li2020neural}. 
Xu et al. employed the physics-informed deep learning by treating the governing equations as a parameterized constraint to reconstruct the missing flow dynamics\cite{xu2021explore}. Wang et al. further applied the physical constraints into the design of neural network, and proposed a grounded in principled physics model: the turbulent-flow network (TF-Net). The architecture of TF-Net contains trainable spectral filters in a coupled model of Reynolds-averaged Navier-Stokes simulation and large eddy simulation, followed by a specialized U-net for prediction. The TF-Net offers the flexibility of the learned representations, and achieves state-of-the-art prediction accuracy \cite{wang2020towards}. 

Most neural network architectures aim to learn the mappings between finite-dimensional Euclidean spaces. They are good at learning a single instance of the governing equation, but they can not generalize well once the given equation parameters or boundary conditions change \cite{raissi2019physics,pan2020physics,wu2020data,xu2021deep}. Li et al. proposed the Fourier neural operators (FNO), which learns an entire family of partial differential equations (PDEs) instead of a single equation \cite{li2020fourier}. The FNO mimics the pseudo-spectral methods \cite{fan2019bcr,kashinath2020enforcing}: it parameterizes the integral kernel in the Fourier space, thus directly learns the mapping from any functional parametric dependence to the solution \cite{li2020fourier}. Benefited from the expressive and efficient architecture, the FNO outperforms the previous state-of-the-art neural network models, including U-Net \cite{chen2019u},TF-Net \cite{wang2020towards} and ResNet \cite{he2016deep}. The FNO achieves 1\% error rate on prediction task of two-dimensional (2D) turbulence at low Reynolds numbers.

Direct numerical simulation (DNS) of the three-dimensional turbulent flows is memory intensive and computational expensive, due to the highly nonlinear characteristics of turbulence associated with the large number of degrees of freedom. In recent years, there has been extensive works dealing with the spatio-temporal reconstruction of two-dimensional turbulent flows \cite{sekar2019fast,chen2019u,cheng2021deep,yousif2021high,hasegawa2020cnn,hasegawa2020machine,chen2021graph,patil2022autoregressive,wen2022u,guibas2021adaptive,pathak2022fourcastnet}. These works reduce the reconstruction error mainly through adopting advanced neural network models \cite{li2020fourier,chen2021graph,peng2022attention,li2022transformer,ye2022learning,li2022fourier,psaros2022meta} or incorporating the prior physical knowledge into the model \cite{raissi2019physics,wang2020towards,kashinath2020enforcing,li2021physics,goswami2022physics,jin2021nsfnets,kashefi2022physics}.

However, modeling of three-dimensional turbulence 
with deep neural networks is more challenging. The size and dimension of simulation data increases dramatically from 2D to 3D \cite{momenifar2022dimension,glaws2020deep}. In addition, modeling the non-linear interactions of such high-dimensional data requires sufficient model complexity and huge number of parameters with hundreds of layers not being uncommon \cite{juefei2017local}. Training such models can be computationally expensive because of the sheer amount of parameters involved. Further, these models also take up a lot of memory which can be a major concern during training, since deep neural networks are typically trained on graphical processing units (GPUs), where the available memory is often constrained.

Arvind et al. first designed and evaluated two NN models for 3D homogeneous isotropic turbulence simulation \cite{mohan2020spatio}. In their work, they proposed two deep learning models: the convolutional generative adversarial network (C-GAN) and the compressed convolutional long-short-term-memory (CC-LSTM) network. They evaluated the reconstruction quality and computational efficiency of the two different approaches. They employed convolutional layers in GANs (CGANs) to handle the high dimensional 3D turbulence data. The proposed CGANs model consists of an eight-layer discriminator and a five-layer generator. The generator takes a latent vector that is sampled from the uniform distribution as an input and produces a cubic snapshot (of the same dimensions as the input) as an output \cite{mohan2020spatio}. The CGANs model has an acceptable accuracy in modelling the velocity features of individual snapshots of the ﬂow, but has difficulties in modelling the probability density functions (PDFs) of the passive scalars advected with the velocity \cite{mohan2020spatio}.

Another model adopts a convolutional LSTM (ConvLSTM) network, which embeds the convolution kernels in a LSTM network to simultaneously model the spatial and temporal features of the turbulence data \cite{shi2015convolutional}. However, the major limitation of ConvLSTM is the huge memory cost due to the complexity of embedding a convolutional kernel in an LSTM and unrolling the network \cite{shi2015convolutional}, especially when dealing with the high-dimensional turbulence data. The authors resolved this challenge of large-size data memory by training the ConvLSTM on the low dimensional representation (LDR) of turbulence data. They  used a convolutional autoencoder (CAE) to learn compressed, low dimensional ‘latent space’ representations for each snapshot of the turbulent ﬂow. The CAE contains multiple convolutional layers, greatly reducing the dimensionality of the data by utilising the convolutional operators \cite{mohan2020spatio}. The convolution filters are chosen since they can capture the complex spatial correlations and also reduce the number of weights due to the parameter-sharing mechanism \cite{lecun2015deep}. The ConvLSTM takes the compressed low dimensional representations as input, and predicts future instantaneous flow in latent space which is then ‘decompressed’ to recover the original dimension \cite{mohan2020spatio}. The CC-LSTM is able to predict the spatio-temporal dynamics of flow: the model can accurately predict the large scale kinetic energy spectra, but diverges in the small scale range.

Nakamura et al. applied the CC-LSTM framework to three-dimensional channel flow prediction task \cite{nakamura2021convolutional}. Despite that the convolutional autoencoder (CAE) can accurately reconstruct the three-dimensional DNS data through the compressed latent space, the LSTM network fails to accurately predict the future instantaneous flow fields \cite{nakamura2021convolutional}. Accurate prediction of three-dimensional turbulence is still one of the most challenging problems for neural networks. 

In recent years, attention mechanism has been widely used in boosting the performance of neural networks on a variety of tasks, ranging from nature language processing to computer vision \cite{vaswani2017attention, parmar2018image,liu2018visual}. The fluid dynamics community has no exception. Wu et al. introduced the self-attention into a convolution auto-encoder to extract temporal feature relationships from high-fidelity numerical solutions \cite{wu2021reduced}. The self-attention module was coupled with the convolutional neural network to enhance the non-local information perception ability of the network and improve the feature extraction ability of the network. They showed that the self-attention based convolutional auto-encoder reduces the prediction error by 42.9\%, compared with the original convolutional auto-encoder \cite{wu2021reduced}. Deo et al. proposed an attention-based convolutional recurrent autoencoder to model the phenomena of wave propagation. They showed that the attention-based sequence-to-sequence network can encode the input sequence and predict for multiple time steps in the future. They also demonstrated that attention based sequence-to-sequence network increases the time-horizon of prediction by five times compared to the plain sequence-to-sequence model \cite{deo2022learning}. Liu et al. used a graph attention neural network to 
simulate the $2D$ ﬂow around a cylinder. They showed that the multi-head attention mechanism can significantly improve the prediction accuracy for dynamic ﬂow fields \cite{liu2022graph}. Kissas et al. coupled the attention mechanism with the neural operators towards learning the partial differential equations task. They demonstrated that the attention mechanism provides more robustness against noisy data and smaller spread of errors over testing data \cite{kissas2022learning}. Peng et al. proposed to model the nonequilibrium feature of turbulence with the self-attention mechanism \cite{peng2022attention}. They coupled the self-attention module with the Fourier neural operator for the $2D$ turbulence simulation task. They reported that the attention mechanism provided 40\% prediction error reduction compared with the original Fourier neural operator model \cite{peng2022attention}. 

The attention mechanism has shown itself to be very successful at boosting the neural networks performance for turbulence simulations, and therefore bringing new opportunities to improve the prediction accuracy of $3D$ turbulence simulation. However, extending the attention mechanism to $3D$ turbulence simulation is a non-trivial task. The challenge comes from the computational expense of the self-attention matrix: the standard self-attention mechanism uses $O(n^2)$ time
and space with respect to input dimension $n$ \cite{vaswani2017attention}. On the other hand, neural networks are often trained on GPUs, where the memory is constrained. Such quadratic complexity has become the main bottleneck for the attention mechanism to be extended to 3D turbulence simulations. Detailed computational cost and memory consumption are discussed in section \ref{sa_module}.

Recently, Wang et al. demonstrated that the self-attention mechanism can be approximated by a low-rank matrix \cite{wang2020linformer}. They proposed the linear attention approximation, which reduces the overall self-attention complexity from $O(n^2)$ to $O(n)$ in both time and space \cite{wang2020linformer}. The linear attention approximation performs on par with standard
self-attention, while being much more memory and time efficient \cite{wang2020linformer}, allowing attention module to be applied on high-dimensional data. In this work, we couple the linear attention module with the Fourier neural operator, for the $3D$ turbulence simulation task.

This paper is organized as follow: section \ref{fno_intro} briefly introduces the Fourier neural operator. Section \ref{attention_intro} compares the standard self-attention with linear attention approximation, and introduces the detailed implementation of coupling attention with Fourier neural operator. Section \ref{dataset} describes the details for generating $3D$ turbulence data. Section \ref{Hyperparamters} discusses the key hyperparamters of the neural network models, including the number of input time steps and the project dimension. In section \ref{benchmark}, we benchmark the prediction performance of the original Fourier neural operator versus the linear attention coupled Fourier neural operator, via statistical and physics-based metrics. In section \ref{discussion} and \ref{conclusion} we provide discussions and draw conclusions respectively.

\section{The Fourier neural operator}\label{fno_intro}

\begin{figure*}
\centering
\includegraphics[width=.7\linewidth]{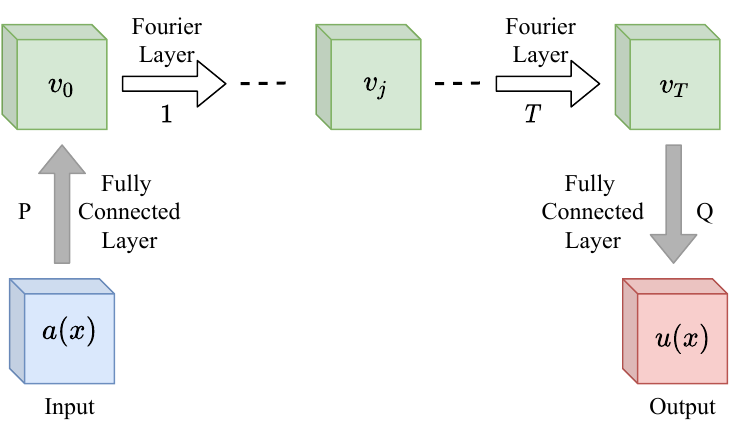}
\caption{Fourier neural operator (FNO) architecture.}
\label{fno_architechture}
\end{figure*}

The Fourier neural operators learn a mapping between two infinite dimensional spaces from a finite collection of observed input-output pairs. Denote $D \subset \mathbb{R}^{d}$ as a bounded open set and $\mathcal{A}=\mathcal{A}\left(D ; \mathbb{R}^{d_a}\right)$,
$\mathcal{U}=\mathcal{U}\left(D ; \mathbb{R}^{d_u}\right)$ as separable Banach spaces of function taking values in $\mathbb{R}^{d_{a}}$ and $\mathbb{R}^{d_{u}}$ respectively \cite{beauzamy2011introduction}. The Fourier neural operators learns an approximation of $\mathcal{A}\rightarrow \mathcal{U}$ by constructing a mapping parameterized by $\theta \in \Theta$. The optimal parameters $\theta^{\dagger} \in \Theta$ are determined in the test-train setting by using a data-driven empirical approximation \cite{vapnik1999overview}. The neural operators are formulated as an iterative architecture $v_{0} \mapsto$ $v_{1} \mapsto \ldots \mapsto v_{T}$ where $v_{j}$ for $j=0,1, \ldots, T-1$ is a sequence of functions each taking values in $\mathbb{R}^{d_{v}}$ \cite{li2020neural}. The FNO architecture, as shown in Fig.\ref{fno_architechture}, consists of three main steps. 

Firstly, the input $a \in \mathcal{A}$ is lifted to a higher dimensional representation $v_{0}(x)=P(a(x))$ by the local transformation $P$. The local transformation $P: \mathbb{R}^{d_a} \rightarrow \mathbb{R}^{d_v}$ acts independently on each spatial component $a(x) \in \mathbb{R}^{d_a}$ of the function $a \in \mathcal{A}$. $P$ is parameterized by a shallow fully connected neural network.
 
Then the higher dimensional representation $v_{0}(x)$ is updated iteratively. In each iteration, the update $v_t \mapsto v_{t+1}$ is defined as the composition of a non-local integral operator $\mathcal{K}$ and a local,
nonlinear activation function $\sigma$. The iteration is described by Eq. (\ref{eq:update}), where $\mathcal{K}:\mathcal{A} \times \Theta_{\mathcal{K}} \rightarrow \mathcal{L}\left(\mathcal{U}\left(D ; \mathbb{R}^{d_{v}}\right), \mathcal{U}\left(D ; \mathbb{R}^{d_{v}}\right)\right)$ maps to bounded linear operators on $\mathcal{U}\left(D ; \mathbb{R}^{d_{v}}\right)$ and is parameterized by $\phi \in \Theta_{\mathcal{K}}$. Here,$W: \mathbb{R}^{d_{v}} \rightarrow \mathbb{R}^{d_{v}}$ is a linear transformation, and $\sigma: \mathbb{R} \rightarrow \mathbb{R}$ is an elementally defined non-linear activation function.

\begin{equation}\label{eq:update}
v_{t+1}(x) =\sigma\left(W v_{t}(x)+\left(\mathcal{K}(a ; \phi) v_{t}\right)(x)\right), \quad \forall x \in D.
\end{equation}

 Lastly, the output $u \in \mathcal{U}$ is obtained by applying the local transformation $u(x)=Q\left(v_{T}(x)\right)$, where $Q: \mathbb{R}^{d_{v}} \rightarrow \mathbb{R}^{d_{u}}$ is parameterized by a fully connected layer. 

Let $\mathcal{F}$ and $\mathcal{F}^{-1}$ denote the Fourier transform and its inverse transform of a function $f: D \rightarrow \mathbb{R}^{d_{v}}$ respectively. Replacing the kernel integral operator in Eq. (\ref{eq:update}) by a convolution operator defined in Fourier space, and applying the convolution theorem, the Fourier integral operator can be expressed by Eq. (\ref{eq:k}), where $R_{\phi}$ is the Fourier transform of a periodic function $\kappa: \bar{D} \rightarrow \mathbb{R}^{d_{v} \times d_{v}}$ parameterized by $\phi \in \Theta_{\mathcal{K}}$. 

\begin{equation}\label{eq:k}
\left(\mathcal{K}(\phi) v_{t}\right)(x)=\mathcal{F}^{-1}\left(R_{\phi} \cdot\left(\mathcal{F} v_{t}\right)\right)(x), \quad \forall x \in D.
\end{equation}

The frequency mode $k \in D$ is assumed to be
periodic, and it allows a Fourier series expansion, which expresses as the discrete modes $k \in \mathbb{Z}^{d}$. The finite-dimensional parameterization is implemented by truncating the Fourier series at a maximal number of modes $k_{\max }=\left|Z_{k_{\max }}\right|=\mid\left\{k \in \mathbb{Z}^{d}:\left|k_{j}\right| \leq k_{\max , j}\right.$, for $\left.j=1, \ldots, d\right\} \mid$. We discretize the domain $D$ with  $n \in \mathbb{N}$ points, where $v_{t} \in \mathbb{R}^{n \times d_{v}}$ and $\mathcal{F}\left(v_{t}\right) \in \mathbb{C}^{n \times d_{v}}$. $R_{\phi}$ is parameterized as complex-valued weight tensor containing a collection of truncated Fourier modes $R_{\phi} \in \mathbb{C}^{k_{\max } \times d_{v} \times d_{v}}$, and $\mathcal{F}\left(v_{t}\right) \in \mathbb{C}^{k_{\max } \times d_{v}}$ is obtained by truncating the higher modes. Therefore $\left(R_{\phi}\cdot\left(\mathcal{F} v_{t}\right)\right)_{k, l}=\sum_{j=1}^{d_{v}} R_{\phi  k, l, j}\left(\mathcal{F} v_{t}\right)_{k, j}, \quad k=1, \ldots, k_{\max }, \quad j=1, \ldots, d_{v}.$

In CFD modeling, the flow is typically uniformly discretized with resolution $s_{1} \times \cdots \times s_{d}=n$, and $\mathcal{F}$ can be replaced by the fast Fourier transform (FFT). For $f \in \mathbb{R}^{n \times d_{v}}, k=\left(k_{1}, \ldots, k_{d}\right) \in \mathbb{Z}_{s_{1}} \times \cdots \times \mathbb{Z}_{s_{d}}$, and $x=\left(x_{1}, \ldots, x_{d}\right) \in D$, the FFT $\hat{\mathcal{F}}$ and its inverse $\hat{\mathcal{F}}^{-1}$ are given by Eq. (\ref{eq:fft}), for $l=1, \ldots, d_{v}$.

\begin{equation}\label{eq:fft}
\begin{aligned}
&(\hat{\mathcal{F}} f)_{l}(k)=\sum_{x_{1}=0}^{s_{1}-1} \cdots \sum_{x_{d}=0}^{s_{d}-1} f_{l}\left(x_{1}, \ldots, x_{d}\right) e^{-2 i \pi \sum_{j=1}^{d} \frac{x_{j} k_{j}}{s_{j}}}. \\
&\left(\hat{\mathcal{F}}^{-1} f\right)_{l}(x)=\sum_{k_{1}=0}^{s_{1}-1} \cdots \sum_{k_{d}=0}^{s_{d}-1} f_{l}\left(k_{1}, \ldots, k_{d}\right) e^{2 i \pi \sum_{j=1}^{d} \frac{x_{j} k_{j}}{s_{j}}}.
\end{aligned}
\end{equation}

\section{Attention enhanced neural network}\label{attention_intro}

\subsection{Self-attention coupled FNO}\label{sa_module}

\begin{figure*}
\centering
\includegraphics[width=.65\textwidth]{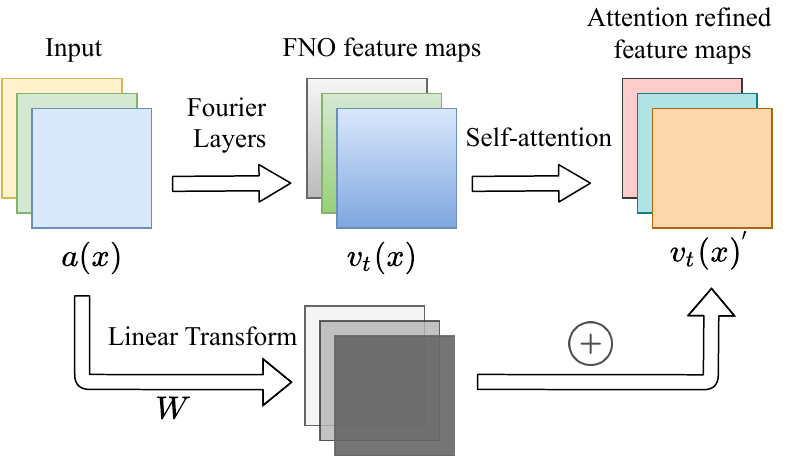}
\caption{Attention enhanced Fourier neural operator.}
\label{fno_attention}
\end{figure*}

\begin{figure*}
\centering
\includegraphics[width=1\textwidth]{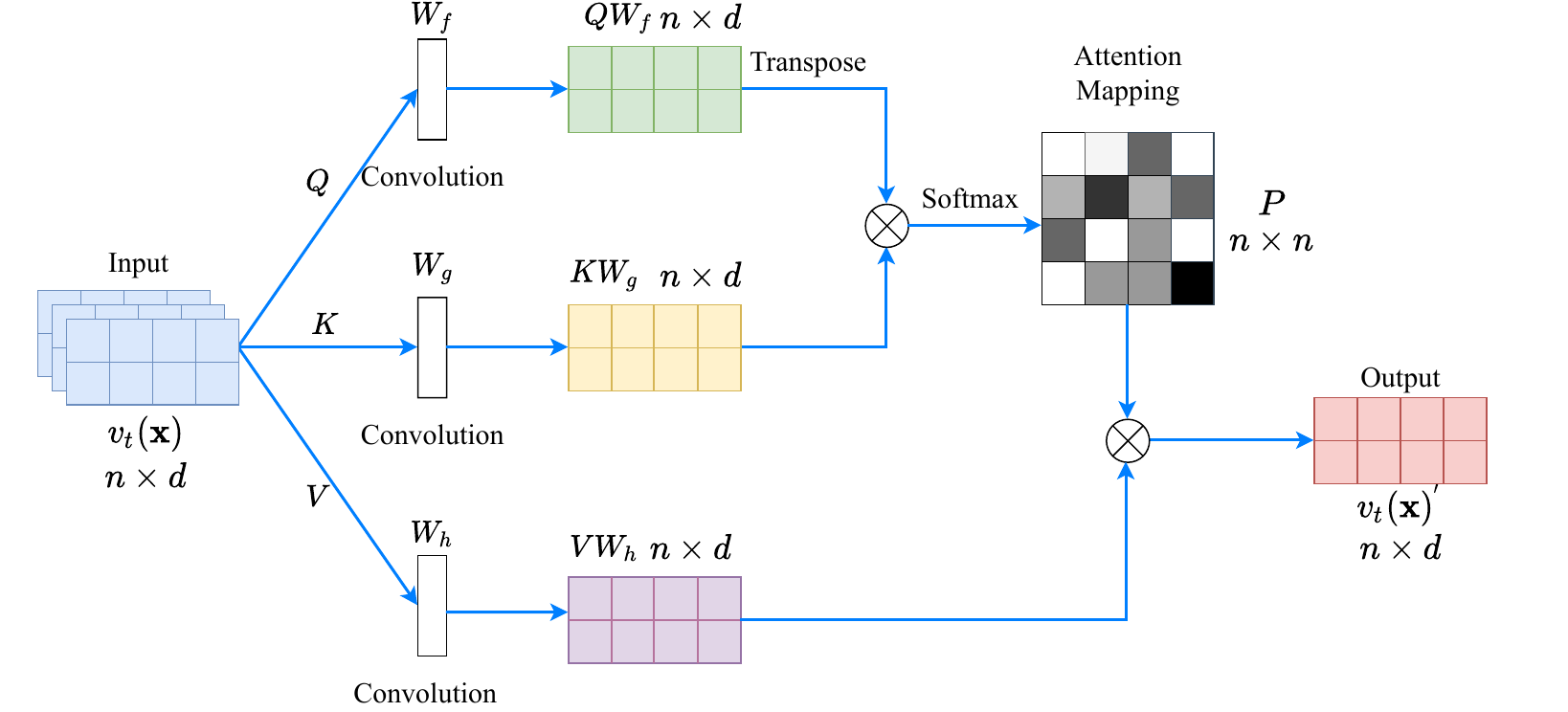}
\caption{Standard self-attention module architecture.}
\label{attention_block}
\end{figure*}

 An attention function can be described as mapping a query and a set of key-value pairs to an output, where the query, keys, values, and output are all vectors. The output is computed as a weighted sum of the values, where the weight assigned to each value is computed by a compatibility function of the query with the corresponding key \cite{vaswani2017attention}. The three sub-modules (query $Q$, key $K$, and value $V$) are the pivotal components of attention mechanism, which come from the concepts of information retrieval systems \cite{kowalski2007information}. Peng et al. proposed to couple the attention mechanism with the Fourier neural operator \cite{peng2022attention}, as shown in Fig. \ref{fno_attention}. The architecture of the self-attention block is shown in Fig.\ref{attention_block}. The convolution parameters $W_{f},W_{g},W_{h}$ learn the embedding of query, key and value, respectively, and these parameters can be jointly learned with the Fourier layers during training. The self-attention block takes the input tensor of shape $n\times d$, and output attention refined feature maps of the same shape, as shown in Eq. (\ref{eq:self-attention}).
 The standard softmax function $\mathbb{R}^K \rightarrow(0,1)^K$ is defined by the Eq. (\ref{softmax}). In the $2D$ turbulence simulation task \cite{peng2022attention}, the $2D$ turbulence data is generated on the grid size of $64 \times 64$, such that $n= 64\times 64$ and $d=20$.

\begin{equation}
v_{t}(\boldsymbol{x})^{'}=\operatorname{Attention}\left(Q W_{f}, K W_{g}, V W_{h}\right)=\underbrace{\operatorname{softmax}\left[\frac{Q W_{f}\left(K W_{g}\right)^{T}}{\sqrt{d}}\right]}_{P} V W_{h}
\label{eq:self-attention}
\end{equation}

\begin{equation}\label{softmax}
\operatorname{softmax}(\mathbf{z})_i=\frac{e^{z_i}}{\sum_{j=1}^K e^{z_j}} \quad \text { for } i=1, \ldots, K \text { and } \mathbf{z}=\left(z_1, \ldots, z_K\right) \in \mathbb{R}^K
\end{equation}
 
 Despite that the self-attention module fits in well with $2D$ turbulence data, however, extending the self-attention module to $3D$ turbulence data becomes challenging and difficult. The main reason is that the standard self-attention operation is computationally expensive in both time and space, especially when dealing with long sequence input \cite{fan2020training,child2019generating,kitaev2020reformer}. It is noted from Fig. \ref{attention_block} and Eq. (\ref{eq:self-attention}) that computing the attention mapping matrix $P$ requires multiplying two $n \times d$ matrices, which incurs a time and space complexity of $O(n^2)$ with respect to input sequence dimension $n$.  In the case of a typical $3D$ flow field of grid size $64\times 64 \times 64 $, the input sequence dimension $m$ becomes $m=64\times 64 \times 64 \times 3$, where the number of physical components is $3$, thus computing the attention mapping requires $2034~GB$ memory for 32-bit floating point data type. Such quadratic complexity on the input sequence dimension has become the main bottleneck for attention to be extended to $3D$ turbulence data simulations. We resolve this bottleneck with the concept of linear attention.
 
\subsection{Linear attention approximation}
 Recently, Wang et al. showed that the  attention mapping matrix $P$  can be approximated by a low-rank matrix $\tilde{P}$ \cite{wang2020linformer}: for any $Q, K, V \in \mathbb{R}^{n \times d}$ and $W_{f}, W_{g}, W_{h} \in \mathbb{R}^{d \times d}$ and for any column vector $w \in \mathbb{R}^{n}$ of matrix $V W_{h}$, there exists a low-rank matrix $\tilde{P} \in \mathbb{R}^{n \times n}$ such that 

\begin{equation}
    \operatorname{Pr}\left(\left\|\tilde{P} w^{T}-P w^{T}\right\|<\epsilon'\left\|P w^{T}\right\|\right)>1-o(1).
\end{equation} 

Here $\operatorname{Pr}$ refers to probability, $\epsilon'$ is a small constant, $o(1)$ is infinitesimal of higher order. Since the attention mapping matrix $P$ is low-rank, it can be approximated by a low-rank matrix $P_{low}$ using the singular value decomposition (SVD) method as below,  where $\sigma_{i}$, $u_{i}$ and $v_{i}$ are the $i$ largest singular values and their corresponding singular vectors.

\begin{equation}
    P \approx P_{\text {low }}=\sum_{i=1}^{k} \sigma_{i} u_{i} v_{i}^{T}=\underbrace{\left[u_{1}, \cdots, u_{k}\right]}_{k} \operatorname{diag}\left\{\sigma_{1}, \cdots, \sigma_{k}\right\}\left[\begin{array}{c}
v_{1} \\
\vdots \\
v_{k}
\end{array}\right]\} k
\end{equation}

However, performing an SVD decomposition in the attention mapping matrix $P$ adds additional complexity. Wang et al. further proposed an efficient approximation method: the linear self-attention \cite{wang2020linformer}. They showed that the attention output can be approximated by adding
two linear projection matrix $E$ and $F$, based on the distributional Johnson–Lindenstrauss lemma \cite{wang2020linformer,johnson1984extensions}.

For any $Q, K, V \in \mathbb{R}^{n \times d}$ and $W_{f}, W_{g}, W_{h} \in \mathbb{R}^{d \times d}$, there exists matrices $E, F \in \mathbb{R}^{n \times k}$ such that \cite{wang2020linformer}, for any row vector $w$ of $Q W_{f}\left(K W_{g}\right)^{T} / \sqrt{d}$, 


\begin{equation}
    \operatorname{Pr}\left(\left\|\operatorname{softmax}\left(w E^{T}\right) F V W_{h}-\operatorname{softmax}(w) V W_{h}\right\| \leq \epsilon'\|\operatorname{softmax}(w)\|\left\|V W_{h}\right\|\right)>1-o(1)
\end{equation}

Detailed mathematical proof can be found in the reference \cite{wang2020linformer}.

\begin{figure*}
\centering
\includegraphics[width=1\textwidth]{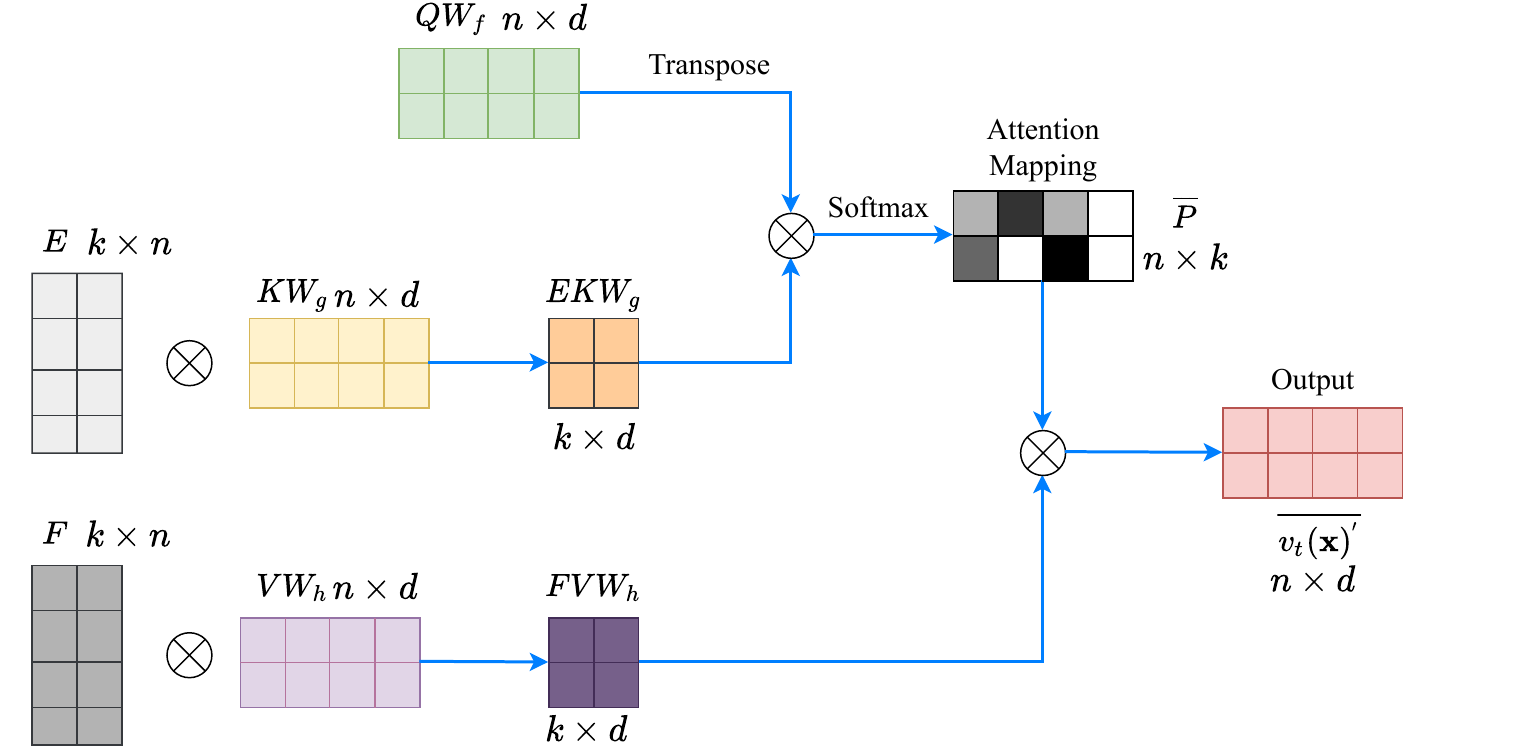}
\caption{Linear self-attention module architecture.}
\label{lin-attention}
\end{figure*}

The architecture of linear attention is shown in Fig. \ref{lin-attention}. It projects the original $(n \times d)$-dimensional learned key and value matrix $KW_{g}$ and $VW_{h}$ into $(k \times d)$-dimensional
projected key and value $EKW_{g}$ and $FVW_{h}$, then computes an $(n × k)$-dimensional attention mapping matrix $\overline{P}$ using the scaled dot-product attention, as described by Eq. (\ref{eq:lin}).  The linear projection matrix $E$ and $F$ are not learnable parameters, instead, they are predefined matrix $E=\delta R$ and $F=e^{-\delta} R$, where $R \in \mathbb{R}^{k \times n}$ with i.i.d. entries from Gaussian normal distribution $N(0,1 / k)$ and $\delta$ is a small constant \cite{wang2020linformer}.

\begin{equation}
\begin{aligned}
\overline{v_{t}(\boldsymbol{x})^{'}} &=\operatorname{Attention}\left(Q W_{f}, EK W_{g}, F V W_{h}\right) \\
&=\underbrace{\operatorname{softmax}\left(\frac{Q W_{f}\left(E K W_{g}\right)^{T}}{\sqrt{d}}\right)}_{\bar{P}: n \times k} \cdot \underbrace{F V W_{h}}_{k \times d},
\end{aligned}
\label{eq:lin}
\end{equation}

\begin{figure*}
\centering
\includegraphics[width=.65\textwidth]{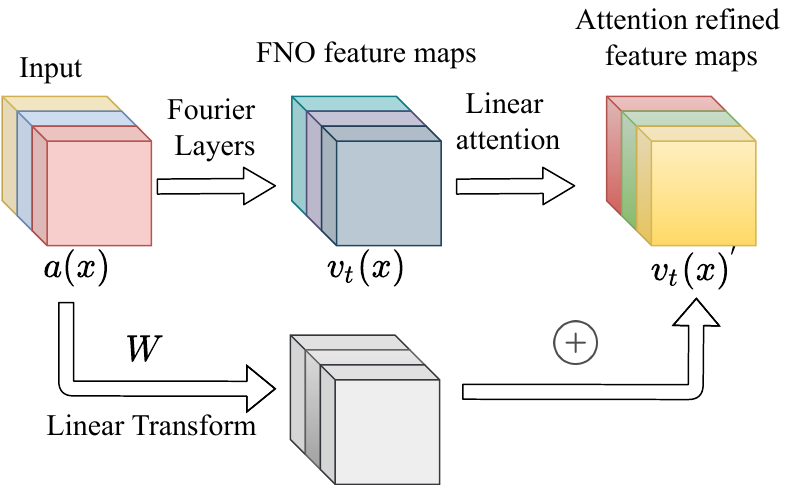}
\caption{Linear attention coupled Fourier neural operator (LAFNO).}
\label{fno_attention_3d}
\end{figure*}

The linear self-attention module performs on par with standard self-attention module \cite{wang2020linformer}, but it offers linear time and memory computation complexity with respect to input sequence length $n$, since the linear self-attention operations only require $O(nk)$ time and space complexity. For a small projected dimension $k$ where $k \ll n$, the memory and space consumption can be significantly reduced, thus allowing the attention mechanism to be applied on high-dimensional data. In this work, we replace the self-attention block by the linear attention module for $3D$ turbulence simulation task. The architecture of our proposed model, the linear attention coupled Fourier neural operator (LAFNO), is shown in Fig. \ref{fno_attention_3d}. Note that the linear transform $W$ in Fig. \ref{fno_attention_3d} is an additional linear transform, different from the linear transform operation inside the Fourier layer \cite{li2020fourier}, to help training the model \cite{wang2017residual}. The linear attention reduces the computational cost of computing the attention mapping matrix $P$, from multiplying two $n \times d$ matrices to multiplying two matrices of $n \times d$ and $k \times d$, where $k \ll n$. 
In our numerical simulation, the maximum GPU memory consumption of training LAFNO is reduced to $35.82~GB$.
 
\section{Dataset description}\label{dataset}

\begin{figure*}
\centering
\includegraphics[width=1\textwidth]{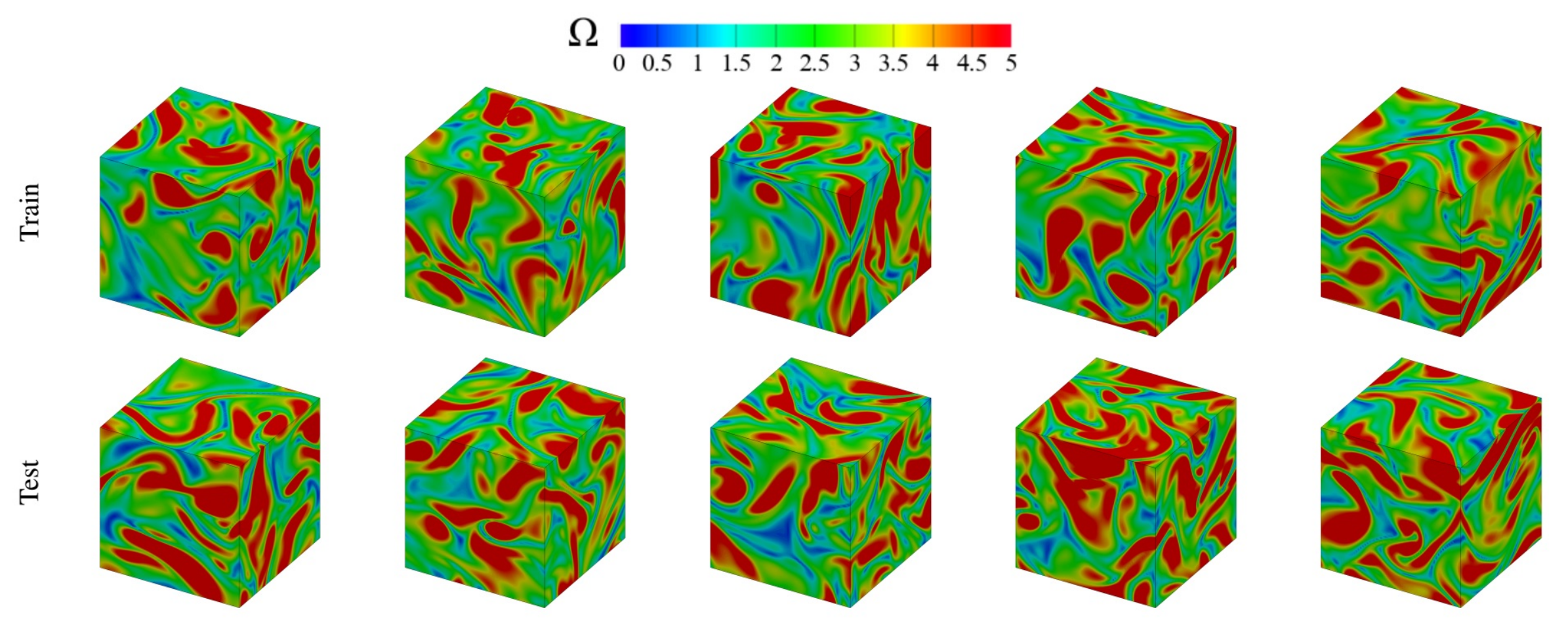}
\caption{Ten random sampled initial fields of vorticity magnitude from the training set and testing set.}
\label{initialization}
\end{figure*}

The dimensionless Navier-Stokes equations in conservative form for the 3D incompressible turbulence are given by \cite{pope2000turbulent}
\begin{equation}
	\nabla  \cdot {\mathbf{u}} = 0,
	\label{ns1}
\end{equation}

\begin{equation}
	\frac{{\partial {\mathbf{u}}}}{{\partial t}} + {\bf{u}} \cdot \nabla {\bf{u}} =  - \nabla p + \frac{1}{{{\mathop{\rm Re}\nolimits} }}{\nabla ^2} {\bf{u}} + \mathbf{\mathrm F},
	\label{ns2}
\end{equation}
where ${\mathbf{u}}$ is the velocity, $p$ is the modified pressure divided by the constant density, ${\mathop{\rm Re}\nolimits}$ is the Reynolds number, and ${\mathbf{\mathrm{F}}}$ is the large-scale force to maintain the turbulence statistically stationary. The initial velocity field ${\mathbf{u}}\left( {t = 0} \right)$ is randomly generated by the Gaussian distribution in spectral space. The initial velocity spectrum of the random velocity field is given by\cite{yuan2021dynamic}
\begin{equation}
	E\left( k \right) = {A_0}{\left( {\frac{k}{{{k_0}}}} \right)^4}\exp \left[ { - 2{{\left( {\frac{k}{{{k_0}}}} \right)}^2}} \right],
	\label{initial_EK}
\end{equation}
where $E\left( k \right)$ is the spectrum of kinetic energy per unit mass, $k$ is the wavenumber magnitude in the spectral space. Here, ${A_0} = 2.7882$ and ${k_0} = 4.5786$\cite{yuan2021dynamic}. The vorticity $\mathbf{\omega}=\nabla \times \mathbf{u}$ measures the local rotation of turbulent eddies, and is a  Galilean-invariant variable used as the inputs and outputs of the neural networks. Fig. \ref{initialization} shows the vorticity magnitude of ten random initial conditions, which are sampled from the training and testing datasets. $\Omega$ is the summation square of vorticity components, as shown in Eq.(\ref{omega}). 

\begin{equation}\label{omega}
    \Omega = \sqrt{\omega_{x}^2+\omega_{y}^2+\omega_{z}^2}
\end{equation}

In this paper, a pseudo-spectral method is applied to numerically simulate the incompressible 3D homogeneous isotropic turbulence in a cubic box of $(2\pi)^3$ on a uniform grid with periodic boundary conditions \cite{wang2020effect,yuan2021dynamic,pope2000turbulent}. The velocity can be expanded as the Fourier series,
\begin{equation}
	{\bf{u}}\left( {{\bf{x}},t} \right) = \sum\limits_{\bf{k}} {{\bf{\hat u}}\left( {{\bf{k}},t} \right){e^{i{\bf{k}} \cdot {\bf{x}}}}} ,
	\label{FourierU}
\end{equation}
where $i$ denotes the imaginary unit, ${i^2} =  - 1$, ${\mathbf{k}}$ represents the wavenumber vector, ${{\bf{\hat u}}}$ is the velocity in Fourier space, and a hat denotes the variable in wavenumber space. The incompressible Navier-Stokes equations in Fourier space are given by \cite{pope2000turbulent}
\begin{equation}
	{\mathbf{k}} \cdot {\mathbf{\hat u}} = 0,
	\label{kns1}
\end{equation}

\begin{equation}
	\left( {\frac{d}{{dt}} + \frac{1}{{{\mathop{\rm Re}\nolimits} }}{k^2}} \right){\mathbf{\hat u}_j}\left( {{\bf{k}},t} \right) =  - i{k_l}\left( {{\delta _{jm}} - \frac{{{k_j}{k_m}}}{{{k^2}}}} \right)\sum\limits_{{\bf{p}} + {\bf{q}} = {\bf{k}}} {{\mathbf{\hat u}_l}\left( {{\bf{p}},t} \right){\mathbf{\hat u}_m}\left( {{\bf{q}},t} \right)}  + {\hat{ \mathcal{F}}_j}\left( {{\bf{k}},t} \right),
	\label{kns2}
\end{equation}
where $\mathbf{p}$ and $\mathbf{q}$ are the wavenumber vectors and  $k_j$ is the $j$-th component of $\mathbf{k}$. The non-local convolution sum at the right-hand side of Eq. (\ref{kns2}) is introduced by the nonlinear convection term and is calculated by the pseudospectral method\cite{yuan2020,pope2000turbulent}. The basic idea is to transform ${\mathbf{{\hat u}_l}}$ and ${{\mathbf{\hat u}_m}}$ to ${{\mathbf{ u}_l}}$ and ${{\mathbf{ u}_m}}$ in physical space by the inverse fast Fourier transform, and then perform the multiplication in physical space, after that use the forward Fourier transform to determine the convolution sum. The non-local convolution sum calculated by the pseudospectral method can greatly reduce the computational cost but will additionally introduce the aliasing errors. The two-thirds rule is used to eliminate the aliasing errors by truncating the Fourier modes with high wavenumbers\cite{yuan2020,yuan2021dynamic,pope2000turbulent}. The large-scale force is constructed by amplifying the velocity field in the wavenumber space to maintain the total kinetic energy spectrum in the first two wavenumber shells to the prescribed values $E_0(1)$ and $E_0(2)$, respectively\cite{yuan2020}.   The forced velocity $\hat u_{j}^f(\mathbf{k})$ is expressed as
\begin{equation}
\hat u_j^f({\bf{k}}) = \alpha {\hat u_j}({\bf{k}}),{\rm{where}}\;\alpha  = \left\{ {\begin{array}{*{20}{c}}
{\sqrt {{E_0}(1)/{E_k}(1)} ,}&{0.5 \le k \le 1.5}\\
{\sqrt {{E_0}(2)/{E_k}(2)} ,}&{1.5 \le k \le 2.5}\\
1&{{\rm{ otherwise}}{\rm{. }}}
\end{array}} \right.
  \label{force_u}
\end{equation}
Here, ${E_0}(1) = 1.242477$ and ${E_0}\left( 2 \right) = 0.391356$\cite{yuan2020}. 

Data are generated on a cubic box of $(2\pi)^3$ with
uniform grid size of $64 \times 64 \times 64$. Time is advanced with the explicit two-step Adams-Bashforth scheme, where the time-step is set to be $\Delta t=0.002$, and the solution is recorded every $t=1$ time units\cite{yuan2021dynamic}. For a partial differential equation ${\partial _t}{{\hat u}_j} = {{\hat R}_j}\left( {{\bf{\hat u}},t} \right)$, the iterative scheme for time advancement is given by Eq. (\ref{scheme_ab2}), where $\Delta t$ is the time step, $t_n = n \Delta t$, and $\hat u_j^n$ denotes the velocity in wavenumber space at time $t_n$. The Taylor Reynolds number of the simulated flow is about 30.  
\begin{equation}
	\hat u_j^{n + 1} = \hat u_j^n + \Delta t\left[ {\frac{3}{2}{{\hat R}_j}\left( {{{{\bf{\hat u}}}^n},{t_n}} \right) - \frac{1}{2}{{\hat R}_j}\left( {{{{\bf{\hat u}}}^{n - 1}},{t_{n - 1}}} \right)} \right].
	\label{scheme_ab2}
\end{equation}

\section{Hyperparameters of neural networks}\label{Hyperparamters}
\subsection{Input time steps T} 

The neural network models take a sequence of previous $T$ time steps of flow field as input. The shape of input tensor is $(T \times H \times W \times D \times C)$, where $H$, $W$ and $D$ is the height, width and depth of 3D flow field respectively, $C$ is the number of vorticity components. Despite that the velocity field can also be used to train the neural networks, our experiments show that the vorticity field provide slightly better testing prediction accuracy, with 2\% error reduction on average. The neural network models predict the flow field of next step as output $(H \times W \times D \times C)$. In the numerical experiments of this paper, $H = W = D = 64$ and $C = 3$. Note that the predicted vorticity at each step is recurrently treated as ground truth and reused as the input (tensors stacked over temporal dimension) with the advance of time. We adopt the standard FNO architecture as described in reference \cite{li2020fourier}. It contains four Fourier layers, the number of truncated Fourier modes is 20, and the number of convolution channels is 36 \cite{li2020fourier}.

The number of input time steps $T$ is one of the key hyperparameters. T was set as 10 for the previous study of 2D turbulence simulations \cite{li2020fourier,peng2022attention}. However, the input number of ten time steps is too expensive for 3D turbulence simulations, since 3D data is memory-intensive. To make trade-off between the prediction accuracy and memory consuming, we conduct experiments with different number of input time steps $T$, ranging from 1 to 10, and track the prediction testing error $\epsilon$ of next step, as shown in table \ref{tab:benchmarkT}. The relative error $\epsilon$ of the model is defined by Eq. (\ref{re_error}), where $\tilde{\omega}$ is the predicted vorticity vector and $\omega$ is the ground truth. 

\begin{equation}\label{re_error}
\epsilon = \frac{\|\tilde{\mathrm{\omega}}-\mathrm{\omega}\|_{2}}{\|\mathrm{\omega}\|_{2}}, 
\text { where }\mathbf{\|a\|}_{2}=1/n
{\sqrt{\sum_{k=1}^{n}\left|{\mathbf{a_{k}}}\right|^{2}}}.
\end{equation}

\begin{table}
    \begin{center}
    \def~{\hphantom{0}}
\begin{ruledtabular}
\begin{tabular}{c|c|c|c|c|c|c|c|c|c|c}
T & 1&2 &3 &4 &5 & 6 & 7 & 8 & 9 & 10 \\ \hline
$\epsilon$
& 0.401 & 0.291 & 0.168 & 0.095 & 0.088 & 0.082& 0.086 & 0.079 & 0.081 & 0.085 \\
\end{tabular}
\end{ruledtabular}
    \caption{Prediction testing errors at different input time steps $T$.}
    \label{tab:benchmarkT}
    \end{center}
\end{table}

We notice from table \ref{tab:benchmarkT} that the error decreases significantly with the number of input steps from $T=1$ to $T=4$, and the error begins to stabilize from $T=4$. Therefore, we choose $T=4$ considering both the prediction accuracy and memory consuming.

We fix the number of input time steps to $T=4$, and investigate the influence of time interval parameter $\delta t$ on the model accuracy, as shown in table \ref{tab:benchmarkt}. The predicting testing error $\epsilon$ is measured at $t=6$, such that prediction interval spans from $t=4$ to $t=6$. We compare four different time intervals $\delta t = 0.2$, 0.4, 1 and 2, corresponding to 100$\Delta t$, 200$\Delta t$, 500$\Delta t$ and 1000 $\Delta t$, respectively, where $\Delta t = 0.002$ is the time interval for DNS that is described in section \ref{dataset}. It is noted that the model performs best around $\delta t=1$. The main reason is that when the time interval $\delta t$ is too small, the temporal accumulation error increases with the increasing of prediction iterations; meanwhile, when the time interval $\delta t$ is too large, it becomes difficult for neural networks to capture the temporal correlations of turbulence at different time steps.


\begin{table}
    \begin{center}
    \def~{\hphantom{0}}
\begin{ruledtabular}
\begin{tabular}{c|c|c|c|c}
$\delta t$ & 0.2 & 0.4 & 1 & 2 \\ \hline
Number of prediction iterations & 10 & 5 & 2 & 1 \\ \hline
$\epsilon$ & 0.129 & 0.122 & 0.106 & 0.133 \\
\end{tabular}
\end{ruledtabular}
    \caption{Prediction testing errors at different time interval  $\delta t$.}
    \label{tab:benchmarkt}
    \end{center}
\end{table}

\subsection{Project dimension k}

Another key hyperparameter is the project dimension $k$ in the linear attention module. $k \ll n$ such that the memory and space can be significantly saved. However, if $k$ is too small, the attention mapping matrix can not be approximated with sufficient accuracy, which further leads to the increase of prediction error. In this work, the $3D$ turbulence data is generated on the grid size of $64\times 64\times 64$, and the number of convolution channels is set as $36$, thus $n=64\times 64 \times 64 \times 3$ and $d=36$. We aim to find the minimum $k$ that provides maximum memory saving while keeping sufficient accuracy. Table \ref{k_table} shows the prediction errors of next step at different $k$. It is noticed that the prediction error starts to increase when $k<36$. The main reason is that the attention mapping matrix $P$ is computed by multiplying two $n \times d$ matrices, where $d \ll n$, thus the attention mapping matrix $P$ is at most rank $d$, and can be approximated accurately for any $k \geq d$.

\begin{table*}
\begin{ruledtabular}
\begin{tabular}{lrcccc}
project dimension $k$ & 
error $\epsilon$ \\
\hline 128 & 0.0956  \\
72 & 0.0964  \\
36 $(k=d)$ & 0.0952 \\
24& 0.107 \\
12 & 0.118 \\
\end{tabular}
\end{ruledtabular}
\caption{Prediction errors at different project dimension $k$.}
\label{k_table}
\end{table*}

\section{Performance benchmark}\label{benchmark}

Since the prediction errors are produced and accumulated at every step, prediction error increases dramatically with time due to the chaotic features of turbulence. In this section, we evaluate the accumulated prediction errors of two models on temporal dimension: the Fourier neural operator (FNO), and the linear attention coupled Fourier neural operator (LAFNO).

\begin{figure*}
\centering
\includegraphics[width=0.6\textwidth]{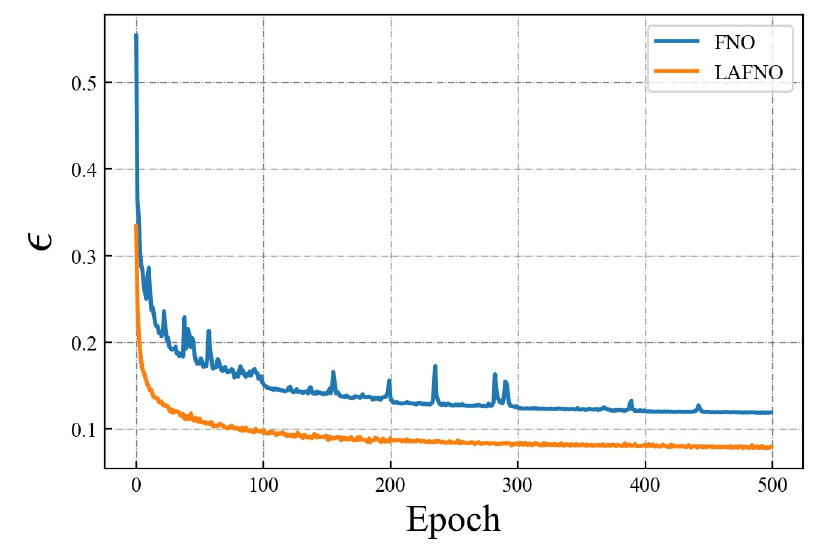}
\caption{Learning curve (testing error) of FNO and LAFNO}
\label{learning_curve}
\end{figure*}

In this numerical experiment, we generate 3000 pairs of input-output data with the numerical solver, where each sample contains 15 steps of solutions of a random initialized condition. The solution is recorded every $t=1$ time units. Both models (FNO and LAFNO) take the vorticity at previous 4 time steps solutions as input, and gives the vorticity at the next time step as output. During training, the vorticity of first 4 steps $\omega|_{(0,2\pi)^{3} \times[1,4]}$ is stacked over temporal dimension as the model input, and the model
recurrently predicts the vorticity at the next step to fit the vorticity at following 11 steps $\omega|_{(0,2\pi)^{3} \times[5,15]}$, which are labeled as the ground truth. Specifically, the predicted vorticity at step 5,  $\hat{\omega}|_{(0,2\pi)^{3} \times[5]}$, is obtained from the input vorticity $\omega|_{(0,2\pi)^{3} \times[1,4]}$. Then the predicted vorticity $\hat{\omega}|_{(0,2\pi)^{3} \times[5]}$ and previous  vorticity sequence $\omega|_{(0,2\pi)^{3} \times[2,4]}$ are stacked over temporal dimension as new input to predict the vorticity at step 6, $\hat{\omega}|_{(0,2\pi)^{3} \times[6]}$, so on and so forth. We use 2400 samples for training and 600 samples for testing. The learning curve (testing error) is shown in Fig.\ref{learning_curve}. It is noted that the LAFNO converges faster and achieves smaller testing error than FNO.  

After training, we evaluate both models on the test dataset, and compare their performance at three selected time steps $t=5$, $10$ and $15$, corresponding to the dimensionless time $t/\tau =5$, $10$ and $15$ respectively. Here $\tau \equiv L_I / u^{r m s} = 0.997$ denotes the large-eddy turnover time \cite{yuan2021dynamic}. $L_{I}$ denotes the integral scale and $u^{rms}$ is the root-mean-square (RMS) of velocity. The interval between every two adjacent time steps is $\Delta T= 1$ and the dimensionless time is $\Delta T/\tau = 1$.

\begin{figure*}
\centering
\includegraphics[width=1\textwidth]{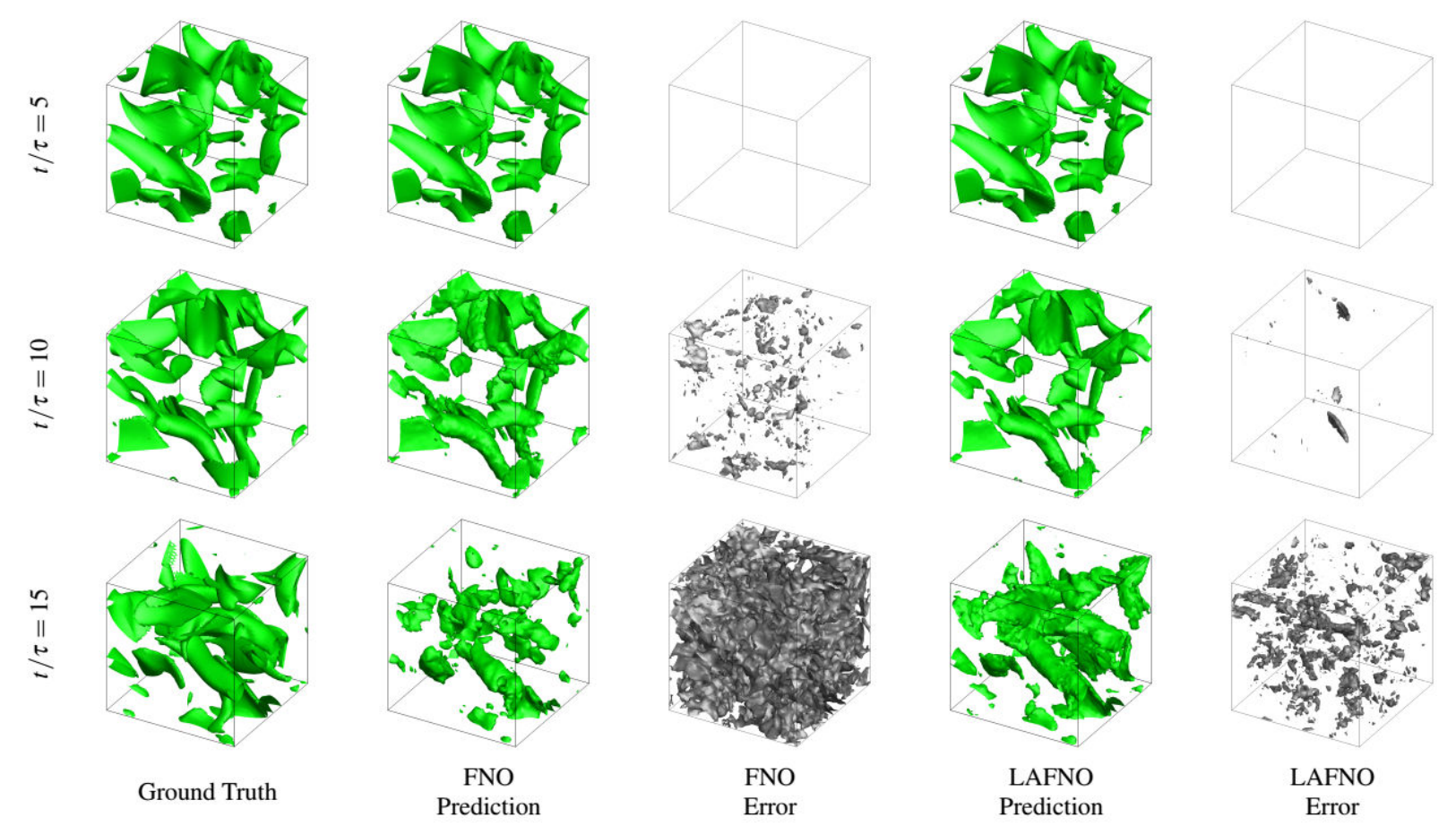}
\caption{Isosurfaces of the normalized vorticity magnitude $\Omega / \Omega^{r m s}=1.5$ and relative error $\epsilon = 0.5$ at selected time steps.}
\label{vor_time_steps_3d}
\end{figure*}

Fig. \ref{vor_time_steps_3d} visualizes the spatial structures of predicted vorticity field and the relative errors of a test sample. Both models can accurately reconstruct the instantaneous spatial structures of turbulence in the beginning. However, the difference is enlarged significantly as time progresses. The FNO error increases significantly with time, in contrast, the errors of LAFNO are visibly smaller in terms of the region.  

\begin{figure*}
\centering
\includegraphics[width=1\textwidth]{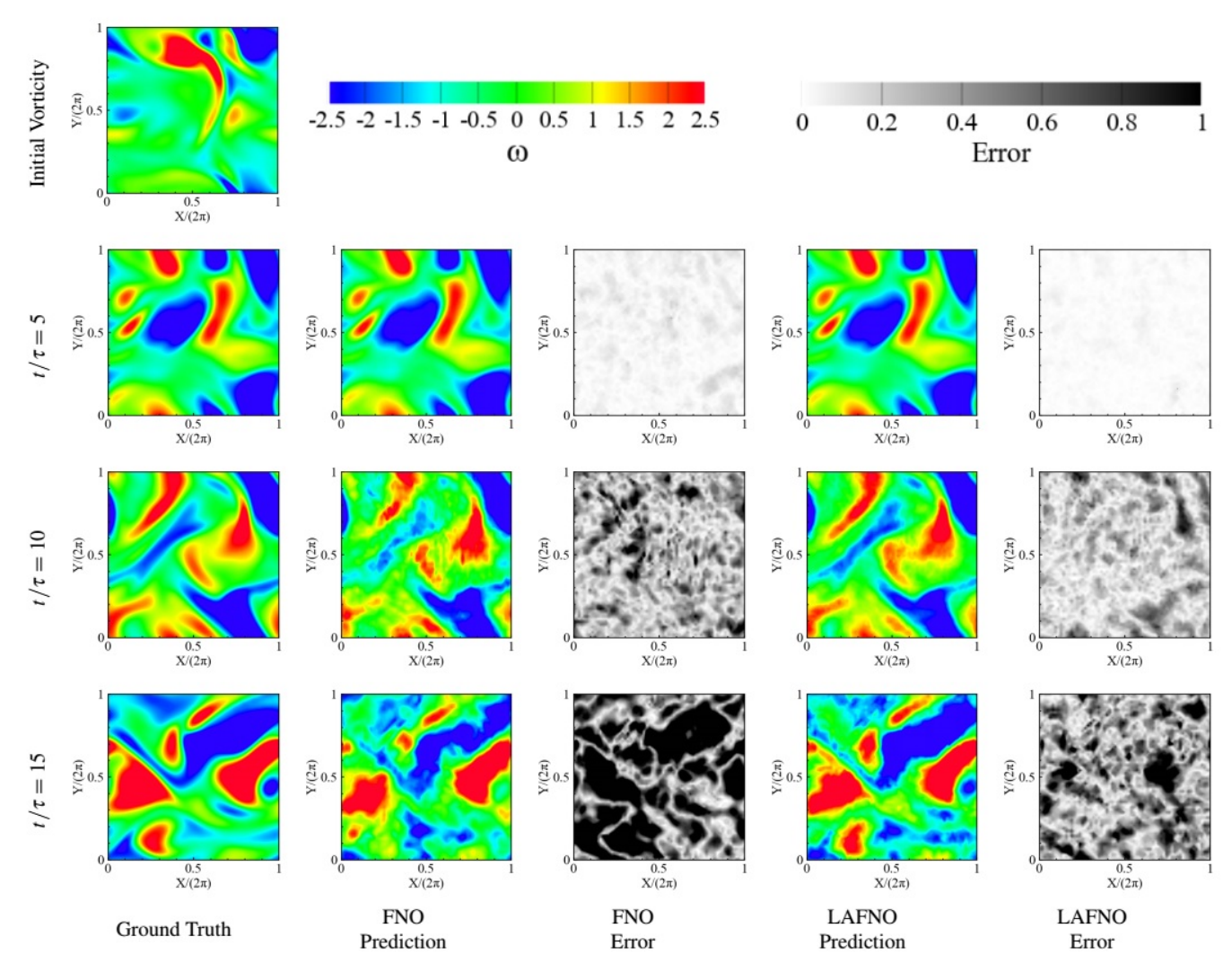}
\caption{Vorticity (2D slice) prediction and absolute error at selected time steps.}
\label{vor_time_steps}
\end{figure*}

Fig. \ref{vor_time_steps} shows a $2D$ vorticity slice in the middle of $Z$ axis from the same test sample in Fig. \ref{vor_time_steps_3d}. At $t/\tau=5$, both models can accurately reconstruct the instantaneous spatial structures of turbulence. As time progresses to $t/\tau=10$, the difference can be visibly noticed: the LAFNO can still make accurate reconstructions whereas the FNO cannot, and LAFNO captures the small-scale structures better than FNO. At $t/\tau=15$, both models fail to reconstruct the instantaneous small-scale structures, whereas the LAFNO can still make relatively accurate reconstructions on the large-scale structures.

\begin{figure*}
\centering
\includegraphics[width=1\textwidth]{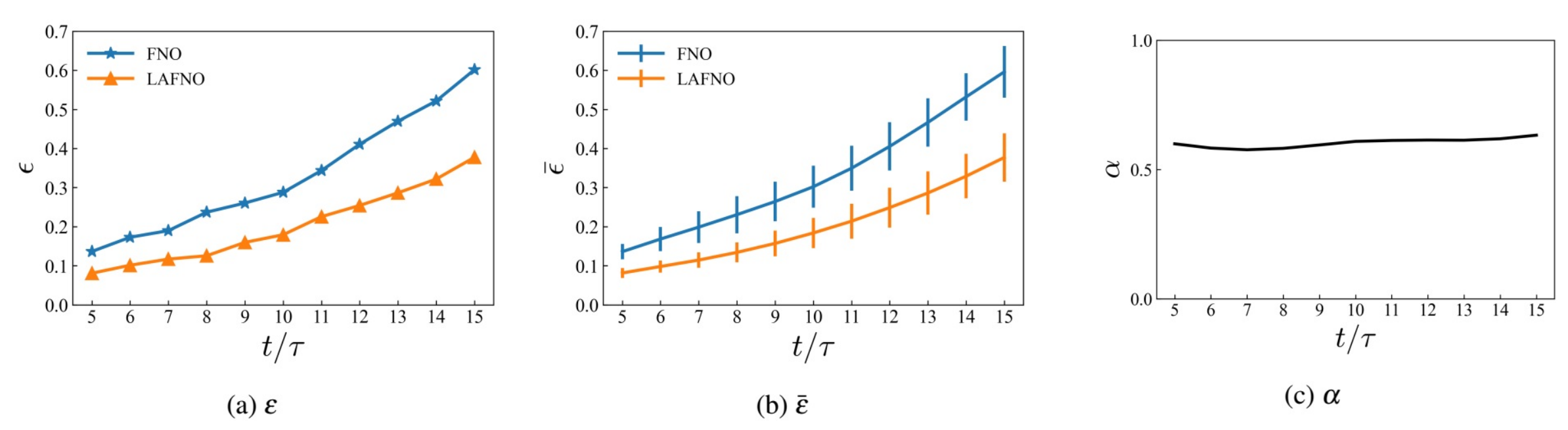}
\caption{Relative error comparison at consequent time steps.~\textbf{(a)} Spatial-averaged relative error of vorticity on single test sample. \textbf{(b)} 
Mean and standard deviation of spatial-averaged relative error of vorticity on 100 test samples. \textbf{(c)} Ratio of mean relative error, where $\alpha = {\bar{\epsilon}}_{LAFNO}/{\bar{\epsilon}}_{FNO}$.}

\label{error_time_steps2}
\end{figure*}

Fig.\ref{error_time_steps2} shows the spatial-averaged relative errors of the two models with respect to consequent time steps. Both models can make accurate predictions in the beginning ($t/\tau=5$) with about 10\% error. However, since the predictions at each step is recurrently treated as ground truth and reused as the inputs with the advance of time, the prediction errors of both models are accumulated iteratively. As time progresses from $t/\tau=5$ to $t/\tau=15$, the error standard deviation becomes larger, and the mean errors of FNO and LAFNO increase to 54\% and 32\% respectively. It is also noticed that the LAFNO achieves 40\% error reduction compared with FNO, throughout all time steps.

\begin{figure*}
\centering
\includegraphics[width= 1\textwidth]{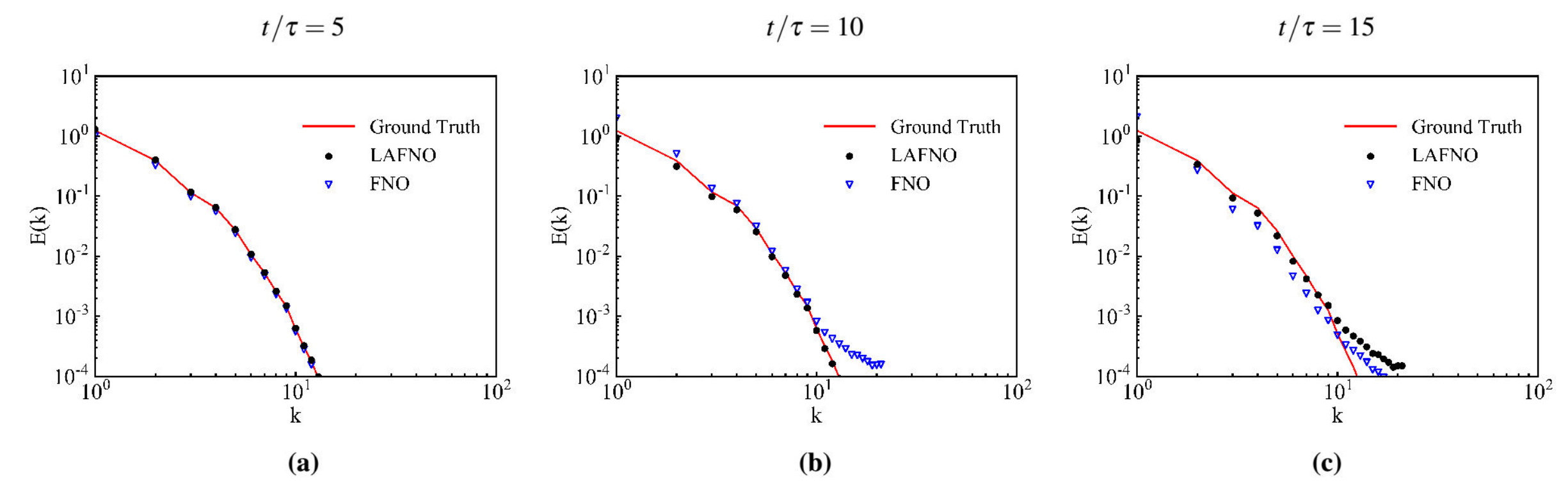}

\caption{Averaged velocity spectrum on 100 test samples at \textbf{(a)} $t/\tau =5$, \textbf{(b)} $t/\tau =10$, \textbf{(c)} $t/\tau =15$.}

\label{spectra_time_steps_avg}
\end{figure*}

Fig.\ref{spectra_time_steps_avg} compares the ensemble-averaged velocity spectrum $\overline{E(k)}$ using 100 test samples. The velocity field is calculated from vorticity field by solving the Poission equation ${\nabla ^2}{\bf{u}} =  - \nabla  \times {\mathbf{\omega}}$. At $t/\tau=5$, the predicted velocity spectrum of both models can agree well with the ground truth in both the low-wave number region and the high-wave number region. As time advances to $t/\tau=10$, the FNO predicted spectrum starts to deviate from the ground truth at the high-wave number region. In contrast, the LAFNO can still accurately capture the small-scale flow structures and well reconstruct the velocity spectrum at different flow scales. At $t/\tau=15$, the predictions of both models deviate from the ground truth at high-wave number region. However, the LAFNO can still make accurate predictions at low-wave number regions whereas the FNO predictions can not.

\begin{figure*}
\centering
\includegraphics[width=1\textwidth]{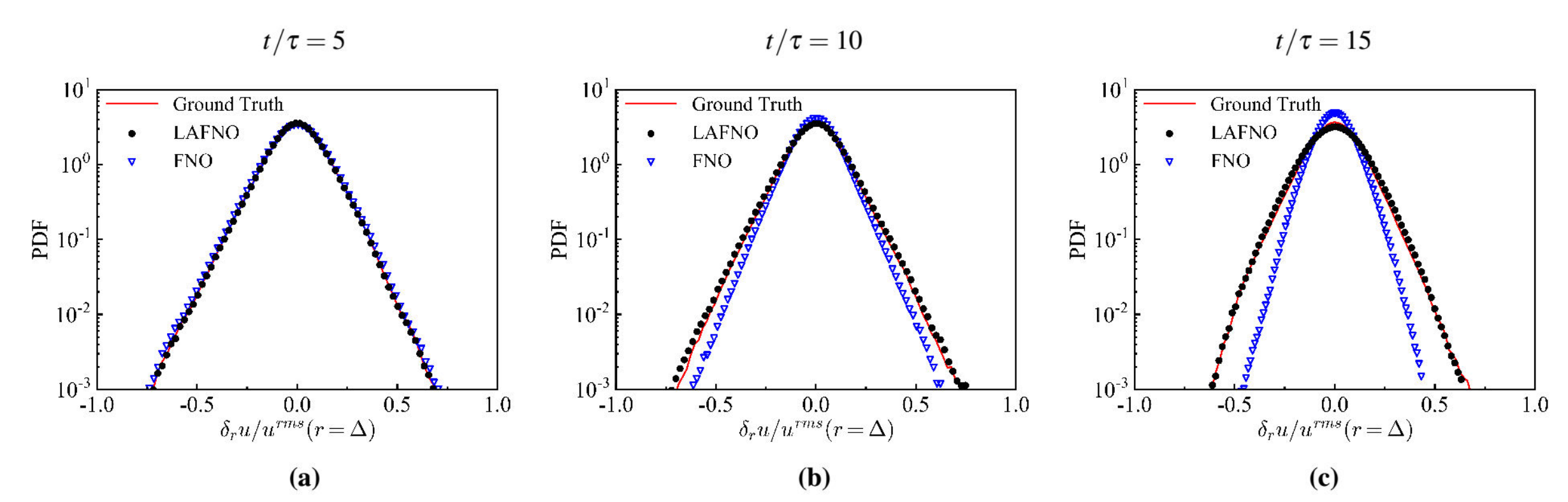}
\caption{PDFs of the normalized velocity increments at different time steps on 100 test samples at \textbf{(a)} $t/\tau =5$, \textbf{(b)} $t/\tau =10$, \textbf{(c)} $t/\tau =15$.}
\label{vel_in_pdf}
\end{figure*}

Fig.\ref{vel_in_pdf} shows the probability density functions (PDFs) of the normalized velocity increment $\delta_{r} {u} / {u}^{r m s}$ at different time steps, where $\delta_{r} {u}=[{\mathbf{u}}(\mathbf{x}+\mathbf{r})-{\mathbf{u}}(\mathbf{x})] \cdot \hat{\mathbf{r}}$ represents the longitudinal increment of the velocity at the separation $\mathbf{r}$. Here, $\hat{\mathbf{r}}=\mathbf{r} /|\mathbf{r}|$, $\Delta$ denotes twice the width of the grid. At the beginning of $t/\tau=5$, the predictions of both models have a good agreement with the ground truth. However, as time advances to $t/\tau=10$ and $t/\tau=15$, the predicted PDFs of FNO become more narrower, whereas the predicted PDFs of LAFNO is always consistent with the ground truth.

\begin{figure*}
\centering
\includegraphics[width= 1\textwidth]{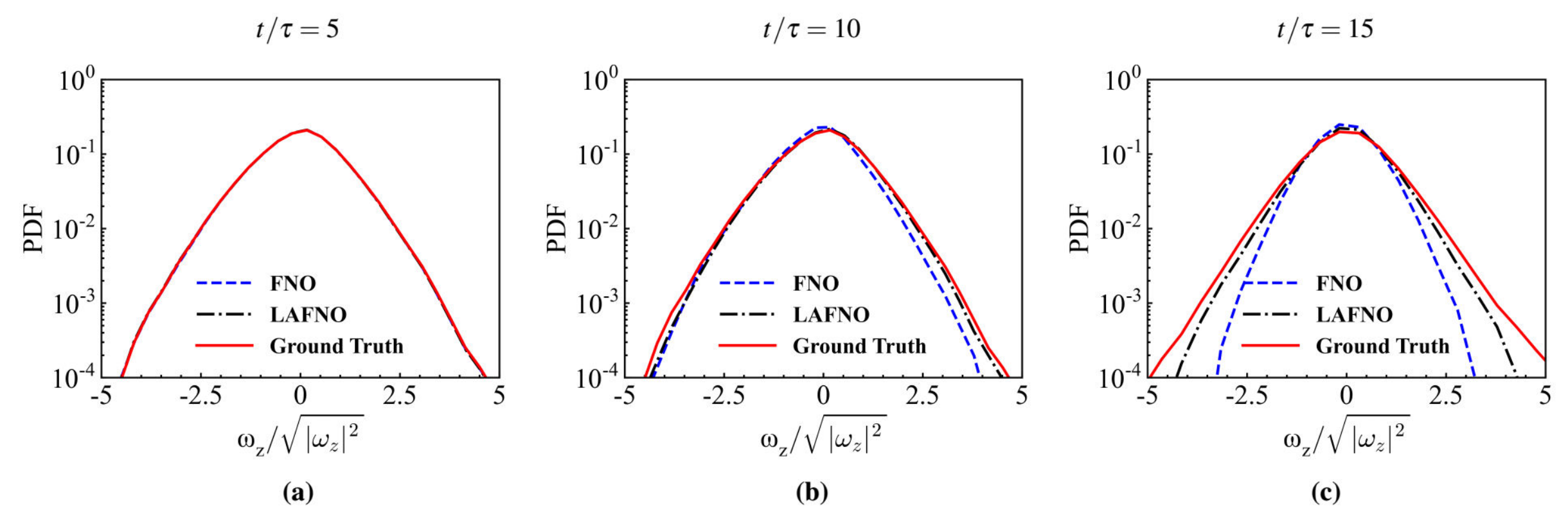}
\caption{PDFs of the normalized vorticity component $\omega_z$ on 100 test samples at \textbf{(a)} $t/\tau =5$, \textbf{(b)} $t/\tau =10$, \textbf{(c)} $t/\tau =15$.}
\label{vor_pdf_time}
\end{figure*}

Fig.\ref{vor_pdf_time} shows the PDFs of vorticity component $\omega_z$ at different time steps. At $t/\tau=5$, the predicted PDFs of both FNO and LAFNO can agree well with the ground truth. At $t/\tau=10$, the predicted PDFs of FNO start to deviate from the ground truth. In contrast, the predicted PDFs of LAFNO can still agree well with the ground truth. As time progresses to $t/\tau=15$, the predicted PDFs of both models deviate from ground truth with LAFNO being significantly closer to the ground truth.

\begin{figure*}
\centering
\includegraphics[width= 1\textwidth]{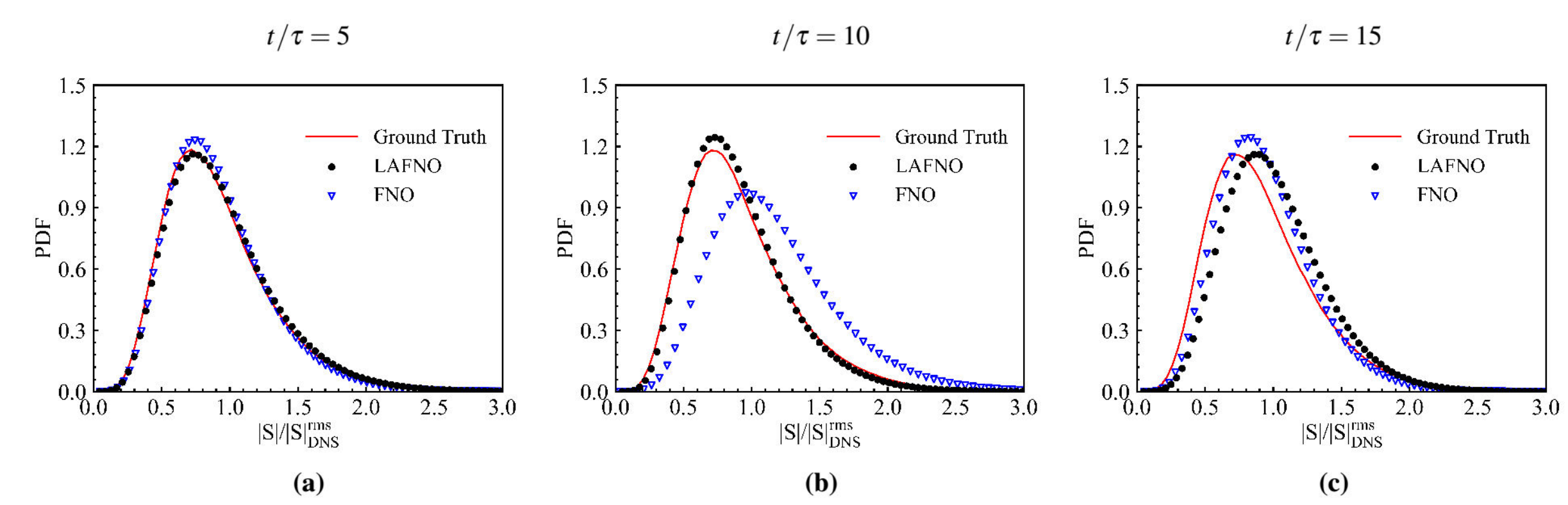}
\caption{PDFs of the normalized characteristic strain-rate on 100 test samples at \textbf{(a)} $t/\tau =5$, \textbf{(b)} $t/\tau =10$, \textbf{(c)} $t/\tau =15$.}
\label{pdf_time}
\end{figure*}

Fig.\ref{pdf_time} illustrates the PDFs of the normalized characteristic strain-rate, namely, $\left| {S} \right|/\left| {S} \right|_{{\rm{DNS}}}^{{\rm{rms}}}$ at different time steps\cite{yuan2021dynamic,yuan2022a}. Here, $\left| {S} \right| = \sqrt {tr\left( {{{{\bf{S}}}^2}} \right)}$ and $\left| {S} \right|_{{\rm{DNS}}}^{{\rm{rms}}} = \sqrt {\left\langle {\left| {S} \right|_{{\rm{DNS}}}^2} \right\rangle }$ are respectively the characteristic strain rate of the predicted velocity field and the root-mean-square values of the characteristic strain rate given by the ground truth, where ``$tr(\cdot)$" denotes the trace of a matrix and ${\bf{S}} = \left[ {\nabla {\bf{u}} + {{\left( {\nabla {\bf{u}}} \right)}^T}} \right]/2$ stands for the strain-rate tensor of the velocity field \cite{yuan2021dynamic,yuan2022a} .  At $t/\tau=5$, the predicted PDFs of LAFNO agree well with the ground truth, while the predicted PDFs of FNO slightly deviates from  the ground truth. As time progresses to $t/\tau=10$ and $t/\tau=15$, the predicted PDFs of both models deviate from ground truth with LAFNO being significantly closer to the ground truth.

\subsection{Generalization on higher Reynolds numbers}

\begin{figure*}
\centering
\includegraphics[width=0.5\textwidth]{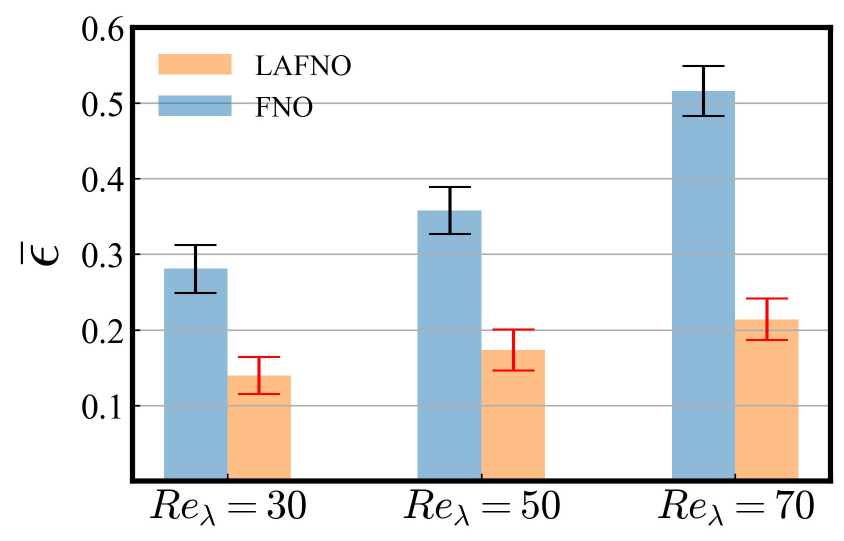}
\caption{Mean and standard
deviation of spatial-averaged relative error of vorticity on 100 test samples at $t/\tau = 10$ for three different Taylor Reynolds numbers.}
\label{Re_gene1}
\end{figure*}

Here, we discuss the generalization performance of FNO and LAFNO on higher Reynolds numbers. We train both models on the dataset described in section \ref{dataset}, at Taylor Reynolds number $Re_{\lambda} = 30$, and evaluate the trained models at $Re_{\lambda} = 50$ and $Re_{\lambda} = 70$. Fig.\ref{Re_gene1} shows 
generalization errors at $t/\tau = 10$ at different Reynolds numbers. It is noted that the prediction errors of both models increase with the increasing of Reynolds numbers. The prediction errors of FNO increase from 0.28 to 0.35 and 0.52, meanwhile, the prediction errors of LAFNO increase from 0.13 to 0.18 and 0.21. The LAFNO performs better than FNO at generalization on higher Reynolds numbers. Moreover, the performance improvement becomes more significant as the Reynolds number gets higher.

\begin{figure*}
\centering
\includegraphics[width=1\textwidth]{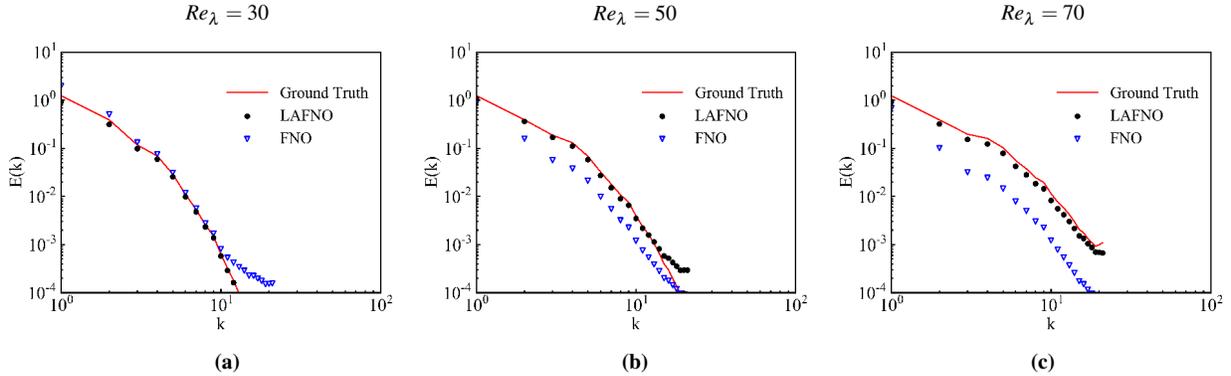}
\caption{Averaged velocity spectrum on 100 test samples at $t/\tau =10$ for Taylor Reynolds numbers of \textbf{(a)} $Re_{\lambda} =30$, \textbf{(b)} $Re_{\lambda} =50$, \textbf{(c)} $Re_{\lambda} =70$.}
\label{Re_gene2}
\end{figure*}

Fig. \ref{Re_gene2} compares the ensemble-averaged velocity spectrum $\overline{E(k)}$  at different Reynolds numbers at $t/\tau=10$ using 100 test samples. At $Re_{\lambda}=30$, the predicted velocity spectrum of LAFNO can agree well with the ground truth in both the low-wave number region and the high-wave number region, meanwhile, the predicted velocity spectrum of FNO deviates from the ground truth at the high-wave number region. As the Reynolds number increases to $Re_{\lambda}=50$, the LAFNO predicted spectrum starts to deviate from the ground truth at the high-wave number region, but can still accurately reconstruct the large-scale flow structures at low-wave number region. In contrast, the FNO predicted spectrum deviates from the ground truth at both the low-wave number region and the high-wave number region. At $Re_{\lambda}=70$, the predictions of both models deviate from the ground truth, with LAFNO being significantly closer to the ground truth.

\subsection{Performance on free shear turbulence}

In addition to 3D homogeneous isotropic turbulence, we also benchmark the performance of FNO and LAFNO on a more complex turbulence simulation task: the 3D free shear turbulence. The 3D free shear turbulence is governed by the same Navier-Stokes equations displayed in Eq. (\ref{ns1} and \ref{ns2}) without the forcing term\cite{wang2022compressibility, wang2022constant}. The simulations of free shear turbulence are performed with lengths ${L_1} \times {L_2} \times {L_3} = 8\pi  \times 8\pi  \times 4\pi $. $x_1$, $x_2$ and $x_3$ respectively denotes the streamwise, normal and spanwise directions, using the uniform grids with ${N_1} \times {N_2} \times {N_3} = 64 \times 64 \times 32$ \cite{wang2022compressibility}. Similar to the 3D homogeneous isotropic turbulence, the periodic boundary conditions in all three directions are adopted and an explicit two-step Adam-Bashforth scheme is selected as the time marching scheme. The Reynolds number is $Re=2000$. The initial conditions of 3D free shear turbulence are given by Eq. \ref{khi_ic} \cite{wang2022compressibility, wang2022constant}.
\begin{equation}\label{khi_ic}
{u_1} = \frac{{\Delta U}}{2}\left[ {\tanh \left( {\frac{{{x_2}}}{{2\delta _\theta ^0}}} \right) - \tanh \left( {\frac{{{x_2} + {L_2}/2}}{{2\delta _\theta ^0}}} \right) - \tanh \left( {\frac{{{x_2} - {L_2}/2}}{{2\delta _\theta ^0}}} \right)} \right] + {\lambda _1},\;{u_2} = {\lambda _2},\;{u_3} = {\lambda _3},
\end{equation}
where $\Delta U$ is the free-stream velocity difference across the shear layer. Here, the magnitudes of perturbation $ \lambda_1 $, $ \lambda_2 $ and $ \lambda_3$ satisfy the Gaussian random distribution where $ \lambda_1, \lambda_2, \lambda_3 \sim \mathcal{N}\left( {0,{{10}^{ - 2}}} \right)$, and ${\delta _\theta ^0=0.08}$  is the initial momentum thickness\cite{wang2022compressibility, wang2022constant}.

We generate 3000 pairs of input-output data with the numerical solver, where each sample contains 15 steps of solutions of a random initialized condition. We use 2400 samples for training and 600 samples for testing. After training, we evaluate both models on the test dataset, and compare their performance at three selected time steps $t=5$, $10$ and $15$, corresponding to the dimensionless time $t/\tau' =125$, $250$ and $375$ respectively, where $\tau'  \equiv \delta _\theta ^0/\Delta U = 0.04$ \cite{wang2022compressibility, wang2022constant,yuan2021dynamic}.
\begin{figure*}
\centering
\includegraphics[width=0.5\textwidth]{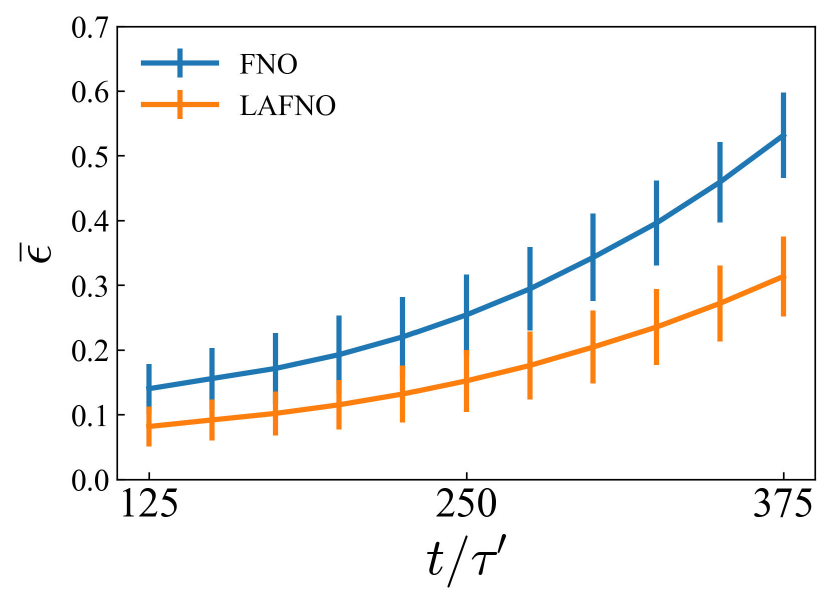}
\caption{Mean and standard
deviation of spatial-averaged relative error of vorticity in free shear turbulence on 100 test samples at different time steps.}
\label{fs1}
\end{figure*}

Fig.\ref{fs1} shows the spatial-averaged relative errors of the two models with respect to consequent time steps. Both models can make accurate predictions in the beginning ($t/\tau'=125$) with about 10\% error. However, since the predictions at each step is recurrently treated as ground truth and reused as the inputs with the advance of time, the prediction errors of both models accumulate iteratively. As time progresses to $t/\tau'=375$, the mean errors of FNO and LAFNO increase to 52\% and 30\% respectively. The LAFNO performs better than FNO throughout all the time steps.

\begin{figure*}
\centering
\includegraphics[width=1\textwidth]{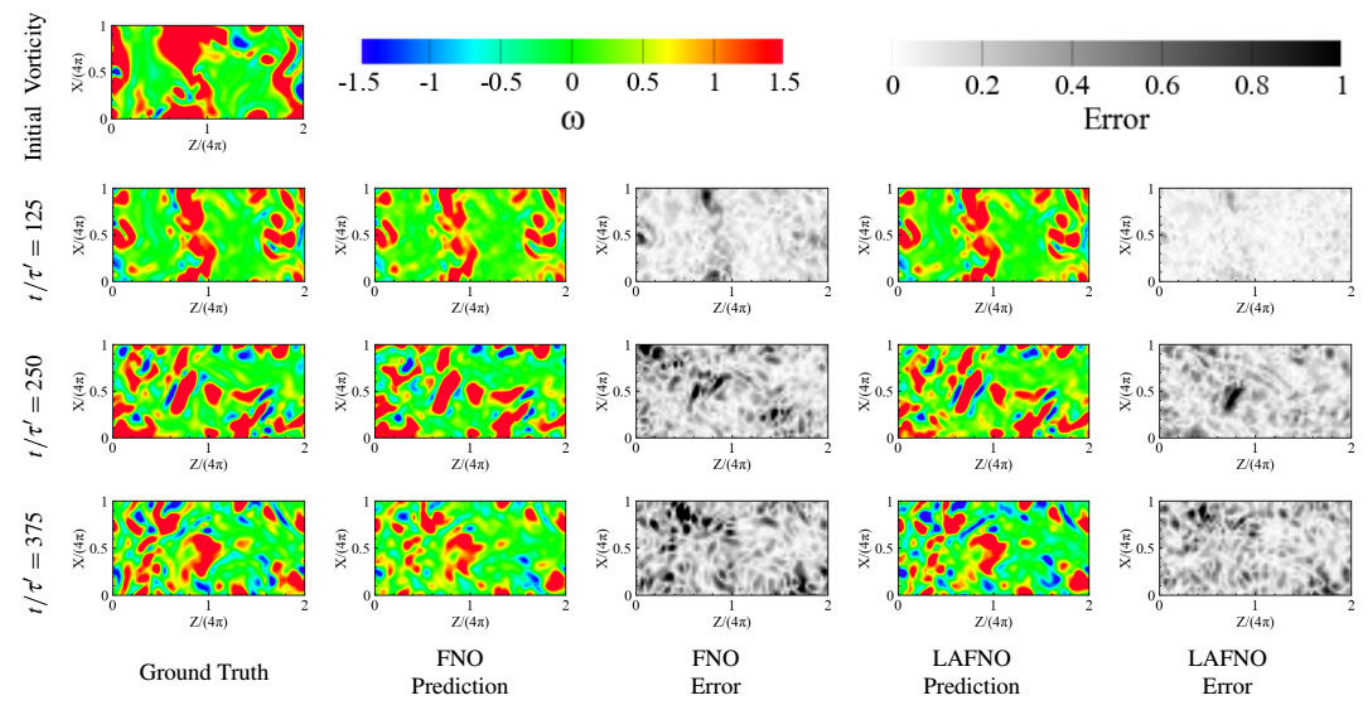}
\caption{Vorticity (2D slice) prediction and absolute error in free shear turbulence at selected time steps.}
\label{fs2}
\end{figure*}

Fig. \ref{fs2} shows a $2D$ vorticity slice in the middle of $Y$ axis from a test sample. At $t/\tau'=125$, both models can accurately reconstruct the instantaneous spatial structures of turbulence. As time progresses to $t/\tau'=250$, the difference can be visibly noticed: the LAFNO can make better reconstructions on the small-scale structures than FNO. At $t/\tau'= 375$, the LAFNO can still make relatively accurate reconstructions on the large-scale instantaneous structures, whereas the FNO cannot. The errors of LAFNO are visibly smaller than FNO in terms of both the magnitude and region.

\begin{figure*}
\centering
\includegraphics[width=1\textwidth]{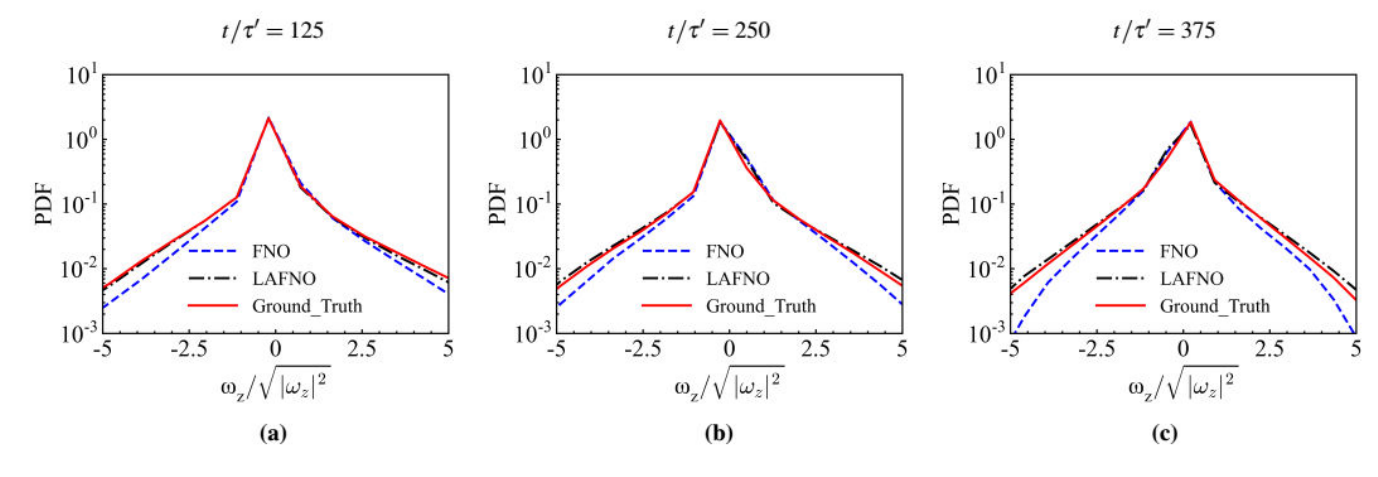}
\caption{PDFs of the normalized vorticity component $\omega_{z}$ in free shear turbulence on 100 test samples at \textbf{(a)} $t/\tau' = 125$,  \textbf{(b)} $t/\tau' = 250$, \textbf{(c)} $t/\tau' = 375$.}
\label{fs3}
\end{figure*}

Fig.\ref{fs3} shows the PDFs of vorticity component $\omega_z$ at different time steps on 100 test samples. At $t/\tau'=125$, the predicted PDFs of LAFNO can agree well with the ground truth, while the the predicted PDFs of FNO deviate from the ground truth. As time progresses to $t/\tau'=250$ and $t/\tau'=375$ , the predicted PDFs of FNO
deviate remarkably from the ground truth, with LAFNO being significantly closer to the ground truth.

\subsection{Computational efficiency}\label{efficiency}

Table \ref{tab:benchmark} compares computational cost of 10 prediction steps on a $64\times64\times64$ grid using 3 different approaches. We implement the numerical experiments on the Pytorch and MindSpore open-source deep learning frameworks. The neural network models (FNO, LAFNO) are trained and tested on Nvidia Tesla A100 GPU, where the CPU type is Intel Xeon(R) Platinum 8350C @ 2.60GHz. The traditional numerical solver is ran on a computing cluster, where the CPU type is 2-sockets Intel Xeon Gold 6148 with 20 cores each @2.40GHz for a total of 40 cores per node. The time consumption of training both models is comparable. Training FNO takes 170 hours, for 500 epochs, where each epoch takes 0.34 hour. On the other hand, training LAFNO takes 168 hours, for 300 epochs, with 0.56 hour per epoch. The LAFNO needs less training epochs than FNO, because LAFNO converges faster, as shown in Fig.  \ref{learning_curve}. Once trained, both models can make efficient predictions to any random initial conditions.  During inference, both neural network models provides 20 folds speedup compared with the DNS approach with the traditional numerical solver. Moreover, the LAFNO achieves 40\% error reduction at the same level of memory consuming and computational expense  as compared to FNO.

\begin{table}[ht]
    \begin{center}
    \def~{\hphantom{0}}
    \begin{ruledtabular}
\begin{tabular}{lccc}
Method & Parameters & Training Time (hrs) & Inference Time (s)\\
\hline
Numerical Solver & N/A & N/A & $12.53s$  \\ 
LAFNO & 70,570,874 & 168 hrs  & $0.576 s$  \\
FNO & 69,605,997 & 170 hrs  & $0.538 s$  \\ 
\end{tabular}
\end{ruledtabular}
    \caption{Computational efficiency comparison of three different numerical approaches.}
    \label{tab:benchmark}
    \end{center}
\end{table}

\section{Discussion and future work}\label{discussion}

Problems in scientific computations and engineering applications often involve solving 3D nonlinear PDEs. However, most existing data-driven approaches have only focused on solving one dimensional (1D) and 2D PDEs, and the 3D problems are rarely discussed and explored. One of the most important reason is that modeling 3D nonlinear PDEs with deep neural networks can be computationally expensive. The size and dimension of simulation data increases dramatically from 2D to 3D. Moreover, modeling such high-dimensional data requires huge number of parameters with hundreds of layers not being uncommon \cite{juefei2017local}. Training and deploying such neural networks can be inefficient when compared with traditional numerical approaches. The FNO has been shown to be one of the most efficient surrogate models in solving PDEs \cite{li2020fourier,li2022fourier,li2022transformer}, thus can be very potential in dealing with 3D nonlinear problems.

Recently, the attention mechanism has been shown to be very promising in boosting the performance of neural networks on solving PDEs \cite{wu2021reduced,deo2022learning,liu2022graph,kissas2022learning}. Peng et al. coupled the self-attention mechanism with FNO, to enhance the FNO prediction ability on $2D$ turbulence simulation \cite{peng2022attention}. The attention mechanism provides 40\% prediction error reduction compared with the original FNO \cite{peng2022attention}. 

However, these works are limited to 2D turbulence simulations, and extending the attention mechanism to 3D turbulence simulation is a non-trivial task. The challenge comes from the computational expense of the self-attention matrix: the standard self-attention mechanism uses $O(n^2)$ time and space with respect to input dimension $n$ \cite{vaswani2017attention}. For a typical $3D$ flow field of grid size $64\times 64 \times 64 $, computing the attention matrix requires $2034~GB$ memory for 32-bit floating point data type. Such prohibitively computational cost has become the main bottleneck for the attention mechanism to be applied on 3D turbulence simulation.

In this work, we explore the possibility of coupling attention with FNO for 3D turbulence simulation. With the linear attention approximation, the memory consumption can be reduced to $35.82~GB$, thus allowing the attention mechanism to be coupled with FNO. Our results show that the linear attention is very effective in boosting the performance of FNO on predicting simple 3D turbulent flows, including 3D homogeneous isotropic turbulence and free shear turbulence. Moreover, the attention mechanism can significantly improve the generalization ability of FNO at higher Reynolds numbers. The proposed linear attention coupled FNO provides an important reference for accelerating 3D complex turbulence simulations.

One limitation of the proposed linear attention is that the problem of error accumulation over time is improved but not fundamentally resolved. In order to solve this problem, physical constraints can be incorporated to ensure that the predictions are subject to the governing equations and conservation laws \cite{raissi2019physics,kashinath2020enforcing,li2021physics,goswami2022physics,jin2021nsfnets,kashefi2022physics}.

Another limitation is that the FNO architecture has only been validated on simple flows including homogeneous isotropic turbulence and free shear turbulence, while the actual engineering flows are usually more complex. Recently, some advanced variants of the FNO architecture have been developed to model complex flows. Improved models including factorized Fourier neural operators (FFNO) \cite{tran2021factorized}, physics-informed neural operator (PINO) \cite{li2021physics}, adaptive fourier neural operators (AFNO) \cite{guibas2021adaptive} and U-shaped neural operators (UNO) \cite{rahman2022u} have been proposed to simulate complex 2D flows.
Li et al. proposed the Geo-FNO \cite{li2022fourier}, to  solve PDEs on arbitrary geometries. The proposed Geo-FNO learns to deform the irregular input domain, into a latent space with a uniform grid, and performs fast Fourier transform on the uniform grid. Such flexibility of handling arbitrary geometries is crucial for solving engineering flows, which usually have more complex geometries with irregular boundaries. However, these advanced FNO variants are still mainly focused on 2D problem, whereas the engineering flows are usually 3D. These FNO variants, including the Geo-FNO, could be extended and coupled with the linear attention and physical constraints to effectively simulate the 3D engineering complex flows in future work.

\section{Conclusion}\label{conclusion}

In this work, we apply the linear attention mechanism to improve neural network models on fast simulation of three-dimensional turbulence. The linear attention approximation reduces the overall self-attention complexity from $O(n^2)$ to $O(n)$ in both time and space, allowing the attention mechanism to be coupled with neural networks and trained on GPUs for 3D turbulence problem. The linear attention coupled Fourier neural operator (LAFNO) is developed and tested for the simulations of 3D turbulence, including homogeneous isotropic turbulence and free shear turbulence.

Numerical experiments show that: 1) the LAFNO can accurately reconstruct a variety of statistics and
instantaneous spatial structures of 3D turbulence.
2) the linear attention can reduce 40\% of the prediction error throughout all the time steps. 
3) the linear attention coupled FNO generalizes better at higher Reynolds numbers than the original FNO.
4) the linear attention coupled FNO model achieves the same level of computational efficiency as compared with the original FNO model. In addition to 3D turbulence simulations, the linear attention can be helpful for the development of advanced neural network models of other 3D nonlinear problems with high-dimensional data.

\section{Data AVAILABILITY}
The data that support the findings of this study are available
from the corresponding author upon reasonable request.
\section*{Acknowledgments}
This work was supported by the National Natural Science Foundation of China (NSFC Grant Nos. 91952104, 92052301, 12172161 and
91752201), by the National
Numerical Windtunnel Project (No.NNW2019ZT1-A04), by the Shenzhen Science and Technology Program (Grants No.KQTD20180411143441009), by Key Special Project for Introduced Talents Team of Southern Marine Science and Engineering Guangdong Laboratory (Guangzhou) (Grant No. GML2019ZD0103), by CAAI-Huawei
MindSpore Open Fund, and by Department of Science and Technology of Guangdong Province (No.2020B1212030001). This work was also
supported by Center for Computational Science and Engineering of Southern University of Science and Technology.

\providecommand{\noopsort}[1]{}\providecommand{\singleletter}[1]{#1}%

\end{document}